\documentclass[]{aa}

\usepackage{graphicx}
\usepackage{txfonts}
\usepackage{natbib}
\bibpunct{(}{)}{;}{a}{}{,} 

\title{Diffuse polarized emission associated with the Perseus cluster}
\titlerunning{Polarization in Perseus}
\author{A.G. de Bruyn \inst{1,2} \and M.A. Brentjens \inst{2,1}}
\authorrunning{de Bruyn \and Brentjens}
\institute{ASTRON, P.O. Box 2, 7990 AA Dwingeloo, the Netherlands \and Kapteyn Astronomical Institute, University of Groningen, P.O. Box 800, 9700 AV, Groningen, the Netherlands}
\offprints{ger@astron.nl}
\date{Received 07 March 2005 / Accepted 8 July 2005}

\abstract{
   We report on full-polarization radio observations of the Perseus
   cluster (\object{Abell~426}) using the Westerbork Synthesis Radio
   Telescope (WSRT) at wavelengths from 81--95~cm. We detect faint,
   very extended polarized emission throughout the cluster region. We
   have employed a novel technique, Rotation Measure synthesis
   \citep{BrentjensDeBruyn2005} to unravel the polarization properties
   of the emission across the full field of view. We detect polarized
   emission over a wide range of RM from about 0 to 90 rad m$^{-2}$.
   Low RM emission (RM $<$ 15 rad m$^{-2}$) is attributed to the local
   Galactic foreground. It has a chaotic structure with smooth changes
   in polarization angle on scales of the order of
   10\arcmin--30\arcmin, not unlike those seen by
   \citet{HaverkornEtAl2003b} at the same frequencies. Emission at
   values of RM $>$ 30 rad m$^{-2}$ on the other hand, shows organized
   structures on scales up to a degree and displays rapidly
   fluctuating polarization angles on scales of the synthesized beam.
   A Galactic foreground interpretation for the high RM emission can
   not be ruled out, but appears extremely implausible. WSRT
   observations at 21~cm of the RM of a dozen discrete sources
   surrounding the Perseus cluster indicate a smooth large-scale
   gradient in the Galactic foreground RM. The diffuse structures have
   a clear excess RM of about 40 rad m$^{-2}$ relative to these
   distant radio galaxies.  This excess Faraday depth, the generally
   good spatial association with the cluster and the different
   morphology of the high RM emission, compared to the genuine
   Galactic foreground emission, all point to an association of the
   high RM emission with the Perseus cluster.  The polarized emission
   reaches typical surface brightness levels of 0.5--1 mJy per
   $2\arcmin\times3\arcmin$ beam and must be rather highly polarized
   ($\ga$ 20\%).  Due to dynamic range limitations and lack of
   sensitivity to large-scale structure we have not yet detected the
   corresponding total intensity. Most of the polarized emission,
   located at distances of about 1$\degr$ from the cluster centre,
   appears too bright, by about 1--2 orders of magnitude, to be
   explainable as Thomson scattered emission of the central radio
   source off the thermal electrons in the cluster. However, this
   remains a viable explanation for the highly polarized 21~cm
   emission from the inner 10\arcmin--20\arcmin and part of the
   81--95~cm emission.  The bulk of the emission associated with the
   Perseus cluster may instead be related to buoyant bubbles of
   relativistic plasma, probably relics from still active or now
   dormant AGN within the cluster. A lenticular shaped structure,
   referred to as the lens, and measuring 0.5--1~Mpc is strikingly
   similar to the structures predicted by \citet{EnsslinEtAl1998}. At
   the western edge of the cluster, we detect very long, linear
   structures that may be related to shocks caused by infall of gas
   into the Perseus cluster along the Perseus-Pisces filamentary
   structure of the cosmic web.

\keywords{Galaxies: clusters: general -- Galaxies: active -- Polarization --
  Magnetic fields  --  Techniques: image processing  -- Radio
  continuum: general  }

}

\begin{document}
\maketitle


\section{Introduction}

\label{brentjens_sec:introduction}

Clusters of galaxies play a crucial role in many fields of
extragalactic research. Located at the crossroads of the filamentary
structures known as the cosmic web, they point to the areas with the
deepest potential wells as predicted in the standard CDM model for
structure formation in the Universe. In addition to the dominant dark
matter and the large numbers of visible galaxies, clusters also
contain enormous reservoirs of diffuse hot gas. This gas is visible
directly through their X-ray emission, and indirectly through their
dynamical effects on galaxies, such as stripping, and the shaping of
head-tail radio sources.

Because of their deep potential wells, clusters continue to accrete
gas from their surroundings. At the cluster-IGM interface this should
lead to (mild) shocks \citep{Burns1998} which might be observed
through enhanced X-ray and radio emission. In the cluster
\object{Abell~3667} infall indeed appears to be responsible for the
generation of very large radio and polarized source structures
\citep{RottgeringEtAl1997, JohnstonHollitt2004}.

The large-scale distribution of the hot gas can in principle be
derived from the X-ray surface brightness profiles. However, there are
two caveats. First, due to the quadratic dependence of X-ray emission
on gas density, this distribution depends on the level of clumpiness
in the medium. Furthermore, temperatures below a few times 10$^6$ K
also lead to reduced X-ray emission and therefore might lead to an
underestimate of the amount of gas. Recent XMM observations
\citep{Kaastra2003} indeed suggest that significant amounts of lower
temperature gas exist at the outskirts of clusters. These observations
may have a direct bearing on the question 'Where are the missing
baryons in the universe?' \citep{CenOstriker1999}. An independent
linear probe of the gas density would be helpful to unambiguously
deduce the gas distribution. The scattering of the 2.7~K microwave
background photons by the cluster electrons, the Syunyaev-Zeldovich
effect, is such a method. Another linear probe of the gas density is
provided through Thomson scattering of radio emission located within
the cluster \citep{Syunyaev1982,WiseSarazin1990}. This method, which
has a fair number of assumptions built in, has a better chance of
detection in low redshifts clusters because of confusion problems.
Until now no unambiguous identification of Thomson scattering in
clusters has been made. Very high dynamic range 21~cm continuum
observations with the WSRT \citep{DeBruynUnpublished1995} may have
detected such emission and were the original motivation for the
observations described in this paper.

Recent radio studies of clusters have drawn attention to another
interesting aspect of clusters, namely the role played by radio loud
AGN in the thermal balance of the intracluster medium. Combined radio
and X-ray images of the inner parts of the Perseus and other rich
clusters with strong radio sources
\citep{FabianEtAl2003ShocksAndRipples,ClarkeEtAl2004} clearly show how the
high pressure radio lobes of the central radio sources are the cause
of cavities in the X-ray surface brightness.

If AGN activity in the radio band is episodic
\citep{SchoenmakersEtAl2000}, clusters should contain many relic, or
fossil, radio sources. Such sources have now been detected in several
clusters
\citep{EnsslinEtAl1998,GovoniEtAl2001,GovoniEtAl2005}. 
\citet{EnsslinGopalKrishna2001} 
show that orphan radio relics would slowly rise in the cluster
atmosphere as a result of buoyancy forces.  The adiabatic energy
losses (on both particles and magnetic field) would then render these
relics effectively invisible at GHz frequencies. These 'bubbles' could
then 'hibernate' for periods of several Gyrs and be invisible except
at extremely low frequencies.  These relics could be 'woken up' by the
shocks associated with the inflow of gas from the surroundings.


In this paper we describe observations of polarized radio emission
from the Perseus cluster which provide a new tool to study the three
components of the diffuse intracluster medium: thermal gas,
relativistic gas, and magnetic fields.  Low frequency radio
polarization data may convey information about extremely low density
regions if they are pervaded by a magnetic field.  Such measurements,
however, are complicated by instrumental and astrophysical effects
which have limited their application in the past. First, off-axis
instrumental polarization of radio telescopes is generally quite
severe at low frequencies.  Using mosaicking techniques to image large
fields will alleviate this to a large degree.  Second, ionospheric
Faraday rotation can rotate polarization angles by up to hundreds of
degrees during the synthesis time at a frequency around 325~MHz. This
can be limited by observing at night and using GPS-based and
ionosonde-based estimates of the ionospheric electron content. Third,
the percentage polarization of radio sources is generally low at low
frequencies as a result of internal depolarization due to Faraday
rotation within the emission regions
\citep{StromConway1985,ConwayStrom1985} and as a result of 
beam depolarization.  
When observing sources with a large rotation measure, or Faraday depth
\citep{Burn1966}, a fourth complication arises: bandwidth depolarization. 

In order for standard polarization analysis (e.g. RM-fitting) to be
meaningful a minimum S/N of about 3 is required in the linearly polarized 
signals for each frequency.  Increasing the bandwidth to improve the
S/N, however, conflicts with the requirement to limit the bandwidth to
avoid bandwidth depolarization.  For example,
observing a polarized source at 350~MHz with a bandwidth of
80~MHz, limits the allowed RM to about 5 rad m$^{-2}$ before Faraday
rotation quenches all polarized signals by an order of magnitude.
To obtain good sensitivity yet avoid bandwidth depolarization 
requires a wideband correlator with a large number of channels.  
As of April 2002 the Westerbork Synthesis Radio Telescope (WSRT) 
can cross correlate, in full polarization, data from all 14 
telescopes in up to 1024 spectral channels over an 80~MHz band.
Employing a technique, which we have called Rotation Measure synthesis 
\citep{BrentjensDeBruyn2005} and which is briefly described below, 
then provides an elegant and powerful method to allow noise limited 
polarimetry using a wide frequency band with 
arbitrarily low S/N ratio in each individual narrow frequency channel.  

A powerful feature of
RM-synthesis is the ability to search for weakly polarized emission
over a wide field of view. A simple version of the method was
developed by \citet{DeBruyn1996RM} for the old WSRT 8 channel
continuum backend.  It was implemented in the WSRT reduction package
NEWSTAR and first applied to data from the highly polarized
millisecond pulsar PSR J0218+4232
\citep{NavarroEtAl1995}. The method will be described briefly in Section
\ref{brentjens_sec:cubes} but a full derivation, discussion 
and analysis of its properties is presented in a companion paper
\citep{BrentjensDeBruyn2005}.

The outline of this paper is as follows. In Section
\ref{brentjens_sec:observations} we present the deep WSRT observations
of the Perseus cluster motivating this paper. Section
\ref{brentjens_sec:reductions} describes the total intensity and
polarization calibration. Section \ref{brentjens_sec:cubes}
describes how the basic data products were made.
Section \ref{brentjens_sec:description} discusses instrumental 
artefacts and how they can be distinguished from
the interesting  astronomical signals.  
Sections \ref{brentjens_sec:foreground} and
\ref{brentjens_sec:background} discuss the nature and location  
of the various types of polarized structures.  Section
\ref{brentjens_sec:discussion} presents our
interpretation of the features that we attribute to 
the Perseus cluster.  Some puzzles and planned future observations are
presented in Section \ref{brentjens_sec:puzzles_and_future}
while Section \ref{brentjens_sec:conclusions_and_summary} concludes.


\section{Observations}

\label{brentjens_sec:observations}

\begin{table*}
\caption{Some parameters of the 92~cm WSRT observations. Dates are formatted
as yyyy/mm/dd hh:mm:ss.} 
\label{tbl:observations}
\begin{center}
\begin{tabular}{lrll}
\hline
Observation ID&9-A (m) & Start date (UTC)& End date (UTC)\\
\hline
10208707 & 72 & 2002/12/02 16:06:20 & 2002/12/03 04:04:50\\
10208767 & 36 & 2002/12/06 15:50:30 & 2002/12/07 03:49:00\\
10208819 & 96 & 2002/12/10 15:34:50 & 2002/12/11 03:33:20\\
10208868 & 84 & 2002/12/12 15:27:00 & 2002/12/13 03:25:30\\
10208937 & 48 & 2002/12/14 15:19:10 & 2002/12/15 03:17:40\\
10208997 & 60 & 2002/12/17 15:07:20 & 2002/12/18 03:05:50\\
\hline
\end{tabular}
\end{center}
\end{table*}

The observations were conducted with the WSRT. 
The array consists of fourteen 25 m dishes on an
east-west baseline and uses earth rotation to fully synthesize the
uv-plane.  For full imaging over the whole primary beam it takes six
array configurations in which the four movable telescopes are stepped
at 12 m increments (i.e. half the dish diameter) with the shortest
spacing running from 36 m to 96 m.  This provides continuous
uv-coverage with interferometer baselines ranging from 36 to 2760 m.
The 12 m increment creates an elliptic grating lobe with a radius in
Right Ascension of about $4\degr$ from the phase centre, which places
it well beyond the 10 dB point of the primary beam which measures
2$\fdg$4 full width at half power. Self confusion is therefore not a 
problem within the $-5$~dB point of the primary beam.  
The angular resolution in Right Ascension is about 0.8$\arcmin$.  
The half power beam width, grating lobe and resolution all correspond 
to a frequency of 350~MHz, in the middle of the observing band, and
scale inversely with frequency. The grating lobe and the
angular resolution furthermore scale with 1/sin($\delta$) in the declination
direction. Basic observational data are collected in
Table~\ref{tbl:observations}. The pointing and phase centre of the
telescope was directed towards (B1950.0): RA =
03$^\mathrm{h}15^\mathrm{m}$, Dec = $+41\degr 15\arcmin$, which is
located between the three dominant radio sources in the Perseus
cluster. This position was chosen to be identical to that of previous
WSRT observations described in
\citet{Sijbring1993} to allow a comparison between the
datasets.

The feed/receiver is part of the new Multi Frequency FrontEnd package
(http://www.astron.nl/wsrt/) and covers the frequency range from about
310--390~MHz. These frequencies correspond to wavelengths from
78--97~cm but for historical reasons, and brevity, we will continue to
refer to this as the 92~cm band.  The 92~cm band is usually largely
free of radio frequency interference (RFI) during evening, nighttime
and weekend observing at the location of the WSRT. The full band can
be completely covered by the new wide band correlator which can process
8 independently tunable bands of 10~MHz.  Each band is covered by 64
channels in 4 cross-correlations to recover all Stokes parameters.  We
used Hamming tapering in the lag-to-frequency transform yielding an
effective spectral resolution of 0.31~MHz. Hamming taper, rather than
Hanning, was used to lower the distant spectral side lobe level 
\citep{Harris1978}. The channel separation was
0.156~MHz. The frequencies of the 8 bands were centred at 319, 328,
337, 346, 355, 365, 374, and 383~MHz. A gap at 360~MHz was introduced
to avoid local RFI. Gibbs ringing at the video edge of the band and
the increased (digital) noise at the upper filtered end of the band
led to useful data in 56 (out of 64) channels of each 10~MHz band
(channels 3--58).  Unfortunately, serious external RFI was encountered
during several hours into the evening at the start of the
observations. In several 10~MHz bands these intense but narrow RFI
signals 'ring' through the whole band, despite our use of a Hamming
taper. By combining pairs of odd-even channels in the subsequent
processing this ringing could be eliminated and the final analysis was
therefore done using 28 channel pairs for each 10~MHz band. Henceforth
we will refer to such a channel pair as a channel. It has an effective
spectral resolution of about 0.4~MHz. The RFI frequently spoiled the
total power monitoring data preventing a precise absolute calibration
of the data. This absolute calibration was therefore done using the
1984 WSRT observations at 327~MHz \citep{Sijbring1993}.

Full syntheses with the WSRT usually need not be interrupted during a
12 hour track.  Because of its equatorial mount -- no change in
parallactic angle -- it also suffices to observe only one polarized
and one unpolarized calibrator to calibrate instrumental leakages.
Every observation was bracketed by two pairs of calibrators, one
polarized and one unpolarized: \object{3C~345} and \object{3C~48} were
observed before the target, while \object{3C~147} and the bright,
highly polarized eastern hot spot in \object{DA~240} were observed at
the end. The calibrator observations lasted 30 min.  The time
resolution in the data was 30~s, which was generally sufficient to
fully sample ionospheric phase fluctuations without serious phase
decorrelation effects. This time sampling was also sufficient to avoid
time smearing for sources at the outer edge of the field.

The system temperature of the WSRT at 350~MHz towards the Perseus
cluster is about 125--150 K. With a net integration time of 72 hours
this should result in a thermal noise of about 20~$\mu\mbox{Jy}$ for
an effective bandwidth of $8\times 8.7$~MHz. However, we did not reach
this level for the following reasons.  Calibration problems did not
allow us to include the upper two frequency bands (374 and 383~MHz).
Due to serious malfunctioning in two telescopes/receivers (RT0 and
RTC), one of which (RTC) provides half of the long baselines in the
array, we decided to use only the inner half of the array in the final
analysis.  This also helped to render the very faint but very extended
polarized emission visible.  To avoid grating lobes, and the inherent
self confusion once emission fills the whole primary beam, we did not
use natural weighting in the imaging (i.e. the redundant baselines
were not included in the imaging process). Hence the final image cubes
were made with only 18 of the maximum 91 baselines for each 12~h
synthesis.  This resulted in a synthesized beam of $2.0\arcmin\times
3.0\arcmin$. Henceforth, when quoting flux densities, we will always
refer to this beam size unless otherwise specified.

The theoretically achievable noise level should increase by a factor
2.5 to 50~$\mu$Jy (this corresponds to 50~mK brightness
temperature). The final noise level, after flagging about 25\% of the
data that were affected by faint RFI or backend problems, was about
70~$\mu$Jy per beam. This, however, is still much lower than the
classical confusion noise of about 1.5~mJy for this frequency and
angular resolution and presents by far the deepest low frequency image
ever made.


\section{Data reduction}

\label{brentjens_sec:reductions}

\subsection{Total intensity calibration}

\begin{figure*}
\centering
\includegraphics[width=\textwidth]{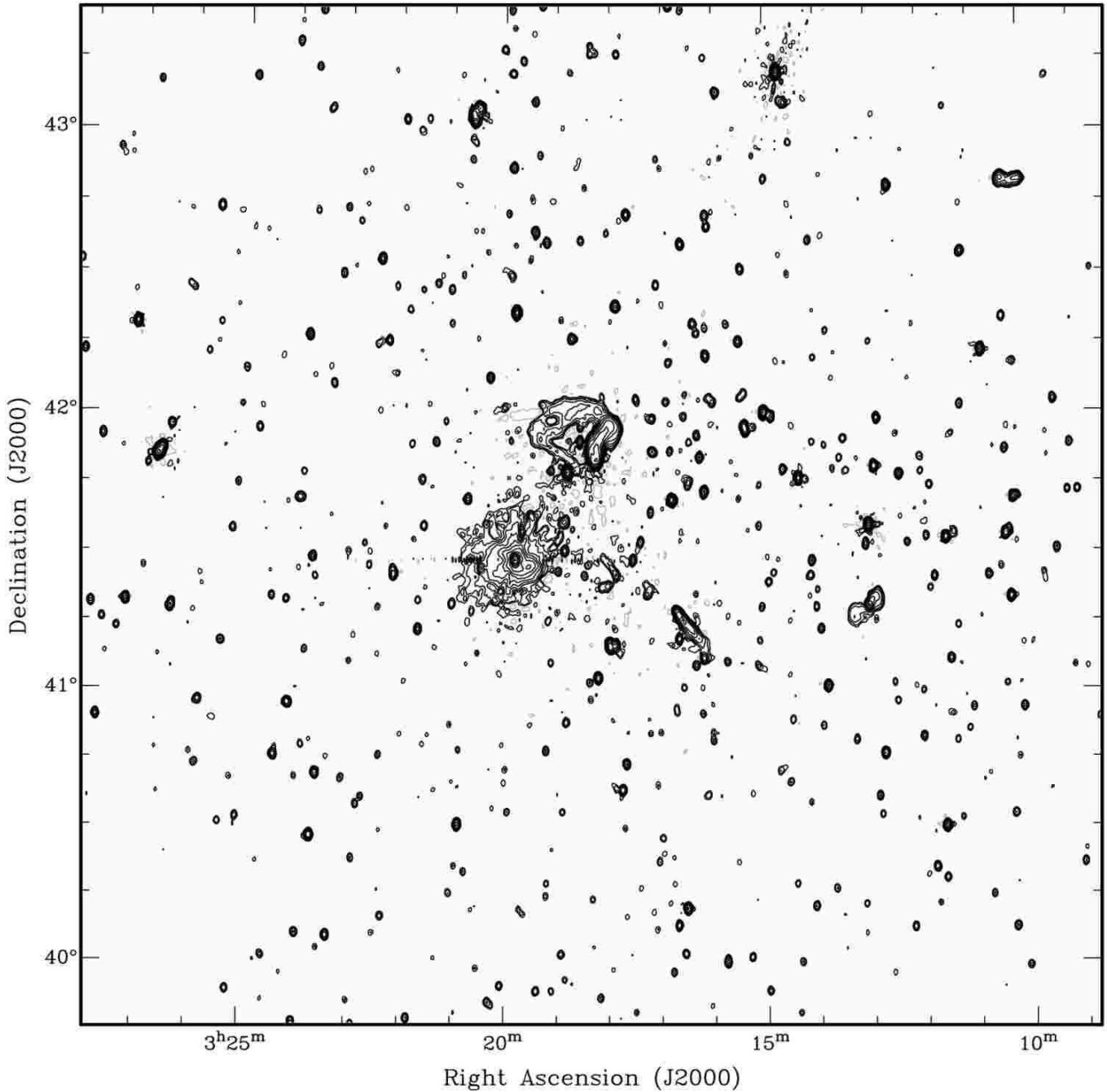}
\caption{Total intensity image of the Perseus cluster, observed with
the WSRT in 2002, using data from band 3. Contour levels are at $-2$, 2,
4, 8, 16, \ldots, 16384 mJy beam$^{-1}$.}
\label{brentjens_fig:total_intensity}
\end{figure*}

The Perseus cluster was first observed with the WSRT at 327~MHz in
1984 \citep{Sijbring1993}. The cluster contains a total of about 50~Jy
of diffuse emission, concentrated in \object{NGC~1275}
(\object{3C~84}) and \object{NGC~1265} (\object{3C~83.1}), and an
unusually rich concentration of bright 4C sources at the edge of the
primary beam (\object{4C~41.08}, \object{4C~42.09},
\object{4C~43.09}).  The complexity of the brightness distribution
necessitated several non-standard steps in the processing which was
done using the WSRT-tailored NEWSTAR reduction package. Initially the
total intensity self-calibration of the data did not converge
satisfactorily. This we attributed to the absence of a sufficiently
accurate initial amplitude calibration (no system temperature
corrections were applied) in the starting model.  This is a well-known
problem in WSRT observations of very complex fields, and is
fundamentally due to the 1-dimensional nature of the WSRT array.  By
using the redundancy constraints in the array \citep{Wieringa1992} we
were able to generate an improved starting image which converged much
better after self-calibration.  Obviously bad data were edited at the
start; more sophisticated flagging was done in an iterative fashion on
the basis of the selfcal residuals.  The total intensity image of the
Perseus cluster is shown in Fig.~\ref{brentjens_fig:total_intensity}.
This image was made from 11 channels in band 3 (average frequency
348~MHz) and is shown at full resolution ($0.9\arcmin\times1.3\arcmin$
beam).  The noise level varies across the image from about 0.3--1.0
mJy~beam$^{-1}$ with the higher values in the inner 2\degr--3\degr of the
imaged area surrounding \object{3C~84} and \object{3C~83.1B}.
Compared to the peak intensity of 18 Jy on \object{3C~84} this
represents a dynamic range of about 20\,000:1.  The observed noise is
only a factor 1.5--2 above the classical confusion noise level.

High dynamic range imaging at low frequency usually is limited by the
phase stability of the ionosphere. In general the ionospheric phase
fluctuations were rather modest during the 6 nights of the
observation, with peak excursions on a 1.4~km baseline of about
$30\degr$ on timescales of 3--5 minutes. In several nights, however,
we experienced periods of a few hours of much faster, temporally
unresolved, scintillation related phase excursions. These phase
variations point to small scale structure in the ionosphere and led to
imaging problems, because they invalidate the standard self-calibration
assumption of position invariant errors. These manifest themselves in
the images as spiky patterns surrounding bright sources at the edge of
the observed field (Note that the WSRT instantaneous response has a
fan beam response which rotates clockwise from position angle
90$\degr$ to 270 $\degr$ during the 12~h synthesis time).  The most
visible ionospheric problems are those surrounding \object{4C~43.09},
which is located about $1\fdg 8$ NNE from the phase centre. They are
also visible around \object{4C~41.08}, located about $1\fdg 6$ east of
the cluster centre.  We also detect weak amplitude errors around these
sources which are due to pointing problems. The results described in
this paper are concerned only with the diffuse polarization in the
inner 3\degr of the cluster and are not affected by pointing.  We have
recently started to experiment with a new iterative 'source-by-source'
self-calibration procedure, called peeling in the context of
LOFAR\footnote{a name suggested by Jan Noordam}, to remove very bright
sources with their own ionospheric phase and telescope complex gain
solutions. Because this paper is mainly concerned with the polarized
emission we will not elaborate on this procedure here any further.


\subsection{Polarization calibration}
\label{brentjens_sec:polarization_calibration}

The WSRT telescopes are equipped with a pair of orthogonal linear
feeds.  All four cross correlations between the incident signal are
formed.  The polarization calibration followed the procedures
described in the papers by \citet{HBS1} and \citet{HBS2}. The overall
(on-axis) instrumental leakages, typically 1--2\%, were calibrated
using the unpolarized calibrator \object{3C~147}. The off-axis
polarization behaviour is much more complicated and is described in
more detail in Section \ref{brentjens_sec:artefacts}.

The instrumental polarization corrections were transferred to the
polarized calibrator source \object{DA~240} which is then used to
phase align the two orthogonal linear polarizations.  The eastern hot
spot of \object{DA~240}, has about 600~mJy of linearly polarized
signal and its RM is about +4~rad~m$^{-2}$. The final polarization
corrections were then transferred to the target source.  The 'peeling'
method alluded to above was also used to correct for the position and
frequency dependent instrumental polarization. This was done on a
channel-by-channel basis. Doing this for 6$\times$28=168 channels made
this a very time consuming process.  For each channel we performed a
full self-calibration as well as polarization calibration using the
same starting model adjusted to take case of the different primary
beam attenuations at the different frequencies. The self-calibration
model included (instrumental) polarization for the brightest discrete
sources.  The 'peeling' was done for 4 different sources
(\object{3C~84}, \object{3C~83.1A} + \object{3C~83.1B},
\object{4C~43.09} and \object{4C~41.08}) in succession. The full
details of this procedure, which are not very relevant for the purpose
of this paper which deals mainly with the polarization results, will
be described elsewhere.

At the observed frequencies ionospheric Faraday rotation can sometimes
be a problem, especially during day-time observing.  We do not have a
strongly polarized source within the Perseus field on which we can
'self calibrate' the ionospheric Faraday rotation.  We therefore made
use of data from two near-simultaneous 12~h syntheses of
\object{NGC~891}, observed for a different project during the 2-week
observing campaign on the Perseus cluster.  This field contains the
highly polarized pulsar \object{PSR~J0218+4232}
\citep{NavarroEtAl1995}.  The pulsar's polarization angle revealed a
Faraday rotation of about 60$\degr$ at the start decreasing to about
zero after 4 hours and remaining small for the remaining 8h of the
observation. The Faraday rotation occurred in the few hours before and
just after sunset.  Because the declination and epoch of the
\object{NGC~891} and Perseus observations were very similar we
adjusted for the 1~h RA-difference and de-applied a polarization angle
rotation of 45$\degr$ decreasing to zero during the first 3 hours of
all six 12~h Perseus cluster syntheses.


\section{From $Q$-$U$ image cubes to RM-cubes} 

\label{brentjens_sec:cubes}

For the final polarization analysis we eliminated 42 channels from the
available 168 because they showed faint traces of RFI or dynamic range
problems related to specific telescopes. The remaining 126 images were
carried through for further processing in RM-synthesis.  The average
noise level in the Stokes $Q$ and $U$ polarization images of a single
channel was rather uniform and varied from about 0.7--0.8~mJy.

The wide range of frequencies in the data set implies that the primary
beam width changes significantly across the band. At the mid-band
frequency of 345~MHz the half power beam width (HPBW) measures about
$1\fdg 22$. The WSRT primary beam is well approximated by a $\cos(c\nu
r)^6$ function where $c$ is a constant equal to about 0.064, $\nu$ is
the frequency in~MHz and $r$ the radius from the pointing centre in
degrees. For the polarized structures discussed in this paper, which
are detected out to a radius of about $1\fdg 5$, the maximum reduction
is about a factor 3.5 at the highest frequency of 370~MHz used in the
analysis. The differential attenuation between 315 and 370~MHz at this
distance from the pointing centre is still only a modest factor 1.4.
We have not tried to correct the data for these differential
effects. In a future paper we will return to the spectral properties
of the polarized emission and, if we manage to recover them, the
associated total intensity structures.


\subsection{Rotation-measure synthesis}

The basic idea behind rotation measure synthesis is that one
derotates, for every pixel in each channel image, the $Q$-$U$ vectors in
order to compensate for a certain assumed rotation measure. 
After derotation, the channel images are averaged. This
procedure maximizes sensitivity to radiation at the assumed Faraday
depth, because that emission is coherently added. All other emission
will add only partly coherently, hence the sensitivity to emission not
at the assumed Faraday depth is reduced. The essence of this procedure
was already mentioned by \citet{Burn1966}. The procedure is not
unknown in the pulsar community
\citep{MitraEtAl2003,WeisbergEtAl2004}. However, it is usually only
applied to a single source or line-of-sight \citep{BowerEtAl1999,
KilleenEtAl1998}. \citet{DeBruyn1996RM} applied the method to every
pixel in WSRT 350~MHz radio synthesis data of the field surrounding
the galaxy \object{NGC~891} and the highly polarized pulsar
\object{PSR~J0218+4232}, but with only 8 frequencies available the
RM side lobes were rather high. Following the completion of the new
broadband (8$\times$10 or 20~MHz) 250\,000 channel WSRT backend the
method was expected to become much more powerful. A full description
of RM-synthesis is presented in a companion paper
\citep{BrentjensDeBruyn2005}. Here we summarize the essential features
of the method as applied to the Perseus cluster data.

The derotation of the multi channel complex polarization images
can be performed very efficiently in terms of
computer time. One first computes the complex polarization $P = Q +
\mathrm{i}U$ and multiplies with a complex phase factor to perform the
rotation. \citet{BrentjensDeBruyn2005} derive that a general inversion
of the polarization as a function of wavelength squared is given by:

\begin{equation}
\{F\ast R\}(\phi) = \frac{\int_{-\infty}^{+\infty}
W(\lambda^2)P(\lambda^2)\mathrm{e}^{-2 \mathrm{i}\phi (\lambda^2 -\lambda_0^2)}
\mathrm{d}\lambda^2}{\int_{-\infty}^{+\infty} W(\lambda^2)\ \mathrm{d}\lambda},
\label{eqn:faraday_dispersion_text}
\end{equation}
where $\phi$ is the Faraday depth,

\begin{equation}
\phi = 0.81 \int_{\mathrm{there}}^{\mathrm{here}}n_\mathrm{e}
\vec{B}\cdot\mathrm{d}\vec{l},
\label{brentjens_eqn:rotation_measure}
\end{equation}
and $F(\phi)$ is the emission as a function of Faraday depth, $\ast$
denotes convolution, $R(\phi)$ the rotation-measure transfer function
(RMTF), $\lambda_0^2$ is the square of the wavelength to which all
vectors are derotated, and $W(\lambda^2)$ the sensitivity as a
function of wavelength squared. This function is also known as the
sampling function, or weight function. $F(\phi)$ is measured in
$\mbox{Jy}\
\mbox{beam}^{-1}\ \mbox{rmtf}^{-1}$. For extended sources that are
discrete in $\phi$, this corresponds to the polarized surface
brightness. For sources that are extended with respect to $\phi$, it
is the polarized surface brightness per RMTF beam width.

If the bandwidth of an individual channel is much less than the total
bandwidth of the observation, we may approximate the weight function
by a sum of $\delta$ functions. That enables us to discretize equation
(\ref{eqn:faraday_dispersion_text}):

\begin{equation}
\{F\ast R\}(\phi_k) = {{\sum_{i=1}^N w_i
P_i\mathrm{e}^{-2\mathrm{i}\phi_k(\lambda_i^2 -
\lambda_0^2)}}\over{\sum_{i=1}^N w_i}}, 
\label{eqn:faraday_dispersion_discrete}
\end{equation}
where $P_i = P(\lambda_i^2)$ for brevity, and $w_i$ is the weight of a
data point. We refer to the companion paper for a formal derivation
of these equations. The RMTF is given by

\begin{equation}
R(\phi) = {{\sum_{i=1}^N w_i
\mathrm{e}^{-2\mathrm{i}\phi(\lambda_i^2-\lambda_0^2)}}\over{\sum_{i=1}^N
w_i}}.
\label{eqn:rmtf_discrete}
\end{equation}

One is free to choose $\lambda_0$. We chose the
wavelength corresponding to the weighted average $\lambda^2$, which
minimizes the component of the transfer function orthogonal to the
actual polarization vector at $\lambda_0$.


\begin{figure}
\resizebox{\hsize}{!}{\includegraphics{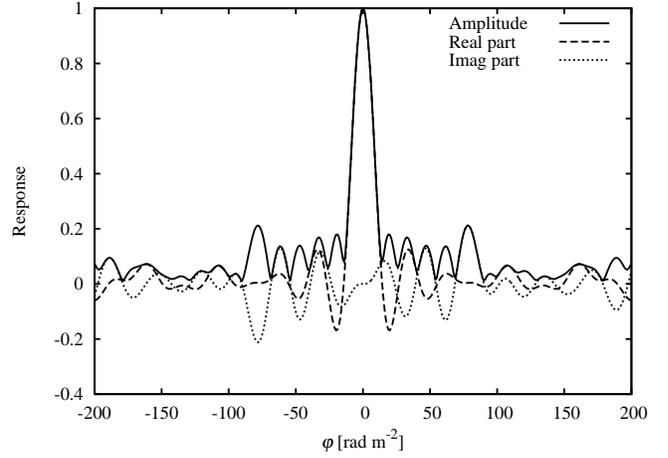}}
\caption{The RMTF of the observations. The FWHM is 15.2 $\mbox{rad}\
\mbox{m}^{-2}$. The real part is the response parallel to the polarization
vector and the imaginary part is the response orthogonal to
it. Because of an uncertainty of the Faraday depth of the peak of
typically a few rad m$^{-2}$ in low S/N cases, no conclusions can be
drawn with respect to the polarization angle at $\lambda = 0$.}
\label{fig:rmtf}
\end{figure}

Fig.~\ref{fig:rmtf} shows the RMTF corresponding to the sampling of
$\lambda^2$ space that was obtained in this work. It shows the
amplitude, the real part, and the imaginary part of the RMTF. The real
part is the response parallel to the polarization vector at
$\lambda_0$ and the imaginary part is the response orthogonal to it.
The output of the RM synthesis procedure is a data cube with axes
$\alpha$, $\delta$, and $\phi$. It has complex values $P = Q +
\mbox{i}U$.


\subsection{The RM-cube}

A total of 126 complex polarization images were used in the
construction of the RM-cube. A range of Faraday depths from $-300$ to
$+300$ rad m$^{-2}$ was synthesized. Beyond this range no features
were detected. The recognition of real astronomical structures as well
as instrumental artefacts is best done by animated scanning through
the cube. Such a movie will be made available on our web page to help
the reader in the recognition of the observed structures.  A selection
of frames from the RM-cube is shown in
Fig.~\ref{brentjens_fig:rmcube}.

RM-synthesis adds polarization images over a considerable range of
frequencies ranging from 315--370~MHz. These images would normally
have angular resolutions that differ by a factor 1.2, or a factor 1.44
in beam area. In generating the half resolution image cube we
therefore tapered to 1/e at a baseline of 1355 wavelengths, providing
an almost frequency independent beam.  In principle we could also
correct the images for the different primary beam
attenuations. However, this would cause a frequency dependent noise
level which depends on radial distance.  A full discussion of the
various issues in dealing with primary beam attenuation, and the
related issue of the intrinsic source emission spectrum, is presented
in \citet{BrentjensDeBruyn2005}.

We have also made RM-cubes with full angular resolution, but we will
only discuss in detail the half resolution images which have better
S/N for the generally extended emission.  We have verified that this
angular smoothing, which locally could lead to some
beam depolarization, did not affect our conclusions in any significant
way.

The addition of 126 complex images, each with a noise level of about
0.7--0.8~mJy, resulted in an RM-cube with rms-noise levels going down
to about 70~$\mu$Jy~beam$^{-1}$ in Stokes $Q$ and $U$. 
However, these levels are only reached at
the edge of the field and/or at high RM values ($>100$~rad~m$^{-2}$)
where instrumental problems have been 'wound up' sufficiently to be
left only with the (uncorrelated) noise in the individual channels.


\section{Description of observed structures}

\label{brentjens_sec:description}

Fig.~\ref{brentjens_fig:rmcube} shows several representative frames
from the total polarization RM-cube. Most of them are separated by
about 9~rad~m$^{-2}$, which should be compared to the resolution in
RM-space of 15~rad~m$^{-2}$.  These images have not been corrected for
the reduced off-axis sensitivity in order to preserve a uniform noise
level across the images.  As discussed above, the images produced
using the RM-synthesis technique utilize an effective bandwidth of
about 50~MHz yielding a sensitivity of better than
100~$\mu$Jy~beam$^{-1}$. This is more than an order of magnitude below
the total intensity noise level and with a peak brightness in Stokes
$I$ of 22 Jy (in the half resolution image) this represents a formal
dynamic range of about 200\,000:1.  It is clear, however, that the
dynamic range varies enormously in the RM-cube. Thus in addition to
the real (astronomical) signals from cluster synchrotron emission, the
polarized images as well as the RM-cube images show many structures
and patterns that have an instrumental origin. We will therefore begin
with a description and, where known, an explanation of these
instrumental artefacts. This will be followed by a description of the
astronomical features.


\subsection{Instrumental artefacts}
\label{brentjens_sec:artefacts}


\begin{figure}
\centering
\resizebox{\hsize}{!}{\includegraphics{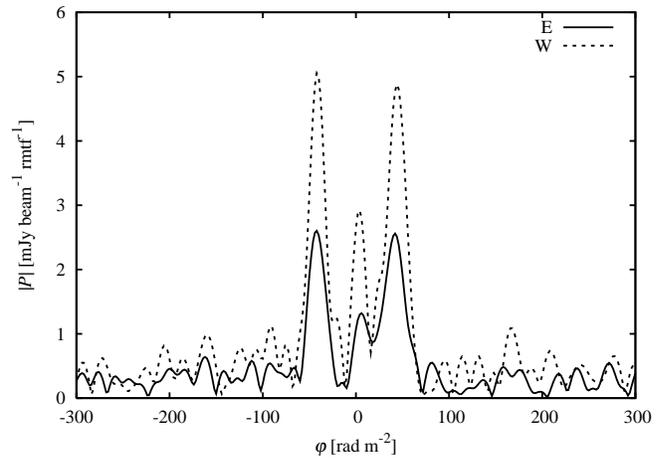}}
\caption{Faraday spectrum at the location of the east (E) and west (W)
  lobe of the extended radio source at $\alpha \approx
  3^\mathrm{h}10^\mathrm{m}, \delta \approx 42\degr 50\arcmin$,
  revealing the instrumental polarization resonances for strong
  off-axis sources.}
\label{brentjens_fig:rm42double}
\end{figure}

We have identified several prominent patterns in the polarization and
RM-cube images that clearly have an instrumental origin.  In
recognizing instrumental artefacts it is important to realize that
they are a combination of uv-plane and image-plane effects.
Multiplicative errors in the uv-plane result in a convolution with an
error pattern in the image plane. The strongest effects are therefore
associated with the strongest sources, although they can extend over a
substantial part of the image plane. The off-axis effects, however,
also have a multiplicative character resulting in a spatially
dependent convolution with an error pattern.  Again the strongest
sources show the largest effects.

Radio emission observed away from the telescope pointing direction of
reflector antennas is generally polarized.  The component that is
uniform across the field of view is small (a few \%) and is calibrated
away using standard procedures (Section
\ref{brentjens_sec:polarization_calibration}). The instrumental
polarization (in Stokes parameters $Q$, $U$ and $V$) in parabolic
dishes, however, rapidly increases with distance from the optical axis
\citep{Napier1999}. In the case of the WSRT it increases to about 20\%
at the $-10$~dB power point with a typical clover-leaf pattern. Stokes
$Q$ is positive in the E-W direction and negative in the N-S
direction. The Stokes $U$ pattern is rotated counterclockwise relative
to Stokes $Q$ by 45$\degr$.
 
\begin{figure}
\centering
\includegraphics[width=\hsize]{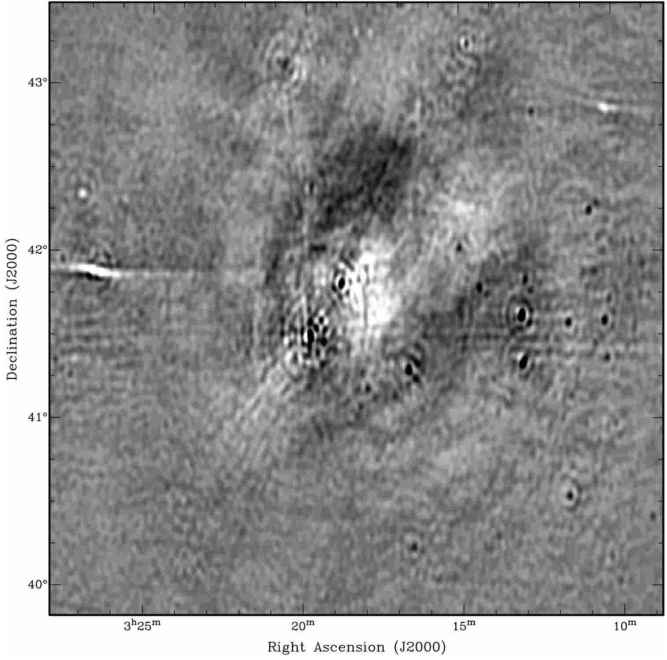}
\caption{Derotated Stokes $Q$ at $\lambda^2 = 0.759\ \mbox{m}^2$
for the foreground emission at a Faraday depth $\phi = 6$ rad
m$^{-2}$. The intensity scale saturates at +1.5 mJy/beam (white) 
and -1.5 mJy/beam (black). }
\label{brentjens_fig:foreground_Q}.
\end{figure}

The off-axis polarization causes spurious signals at the location of
the many strong sources in the field (see
Fig.~\ref{brentjens_fig:foreground_Q}). The inevitable small spread in
off-axis polarization levels for different telescopes in the array
also causes weak ring-like patterns (see
Fig.~\ref{brentjens_fig:foreground_Q}).  Furthermore, the small
variations between different 12~h observations also cause weak
residuals at multiples of the grating lobes radius (about 40$\arcmin$,
and frequency dependent). Intersection of these grating lobes then
gives rise to a number of weak spurious features.

At RM=0, all contributions add up coherently, thus if the off-axis
polarization would be independent of frequency the response in the
RM-cube should rapidly diminish at large RM values.  The observed
response as a function of RM indeed drops fairly rapidly but not as
fast as expected.  This is due to the fact that the off-axis
polarization of the WSRT, in Stokes $U$ and to a lesser extent Stokes
$Q$, also reveals a very strong frequency dependence with a period of
about 17~MHz.\footnote{This instrumental attribute has been known to
exist for a long time but its origin is still not fully understood. It
is believed that there is a component due to a standing wave pattern
between the focus box and the dish (separated by 8.75~m) and a
component due to scattering off the four legged support structure of
the focus box.}


The 17~MHz period causes peaks in the RM 'spectrum' at values of about
$+42\ \mbox{rad}\ \mbox{m}^{-2}$ and $-42\ \mbox{rad}\ \mbox{m}^{-2}$,
in addition to the emission at $0\ \mbox{rad}\ \mbox{m}^{-2}$. These
'resonances' can be clearly discerned in intense sources located far
off-axis when scanning through the RM movie.  The elongated source in
the north-western upper corner of the image
(Fig.~\ref{brentjens_fig:total_intensity}) shows this very clearly
(Fig.~\ref{brentjens_fig:rm42double}). Detailed scrutiny of the
Faraday spectrum over the face of the source suggests that the part of
the emission around $\phi$ is +5 rad m$^{-2}$ may in fact be partly
due to intrinsic polarization from the source. The predicted Galactic
value at this location (see Section \ref{brentjens_sec:background},
Fig.~\ref{brentjens_fig:rm_maxmap}) is about $-10$~rad~m$^{-2}$.



\begin{figure}
\centering
\includegraphics[width=\hsize]{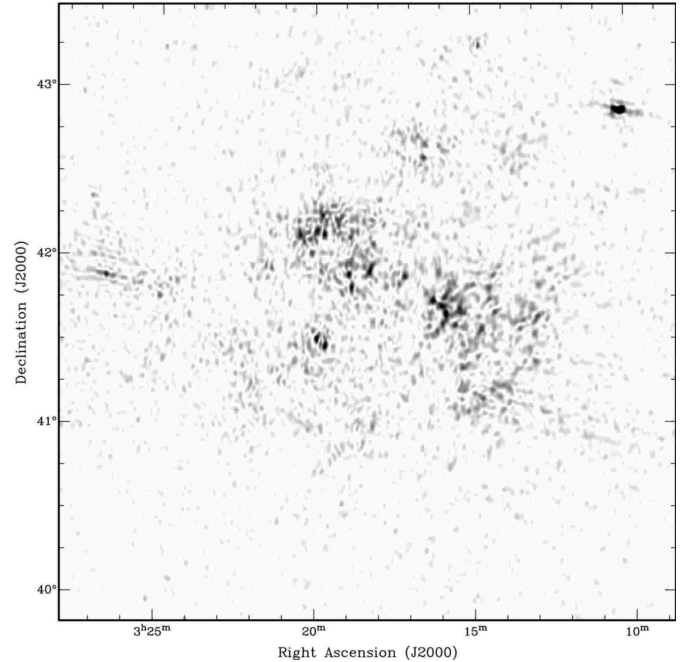}
\caption{Polarized intensity $\|P\|$ at $\phi$= 52~rad~m$^{-2}$.
The grey scale covers a range from 0.3--2~mJy~beam$^{-1}$~rmtf$^{-1}$.}
\label{brentjens_fig:doughnut_lens}
\end{figure}

A third instrumental artefact in the images is a faint pattern of
radial stripes emanating from \object{3C~84}, by far the strongest
radio source in the field. The intensity of these spikes rapidly dies
out when we move away from position angles of $\pm 90\degr$, giving
them the appearance of 'whiskers'. Their peak intensity is about
0.5~mJy~beam$^{-1}$ but they are more typically present at
0.1--0.2~mJy~beam$^{-1}$.  These whiskers are best seen in
Fig.~\ref{brentjens_fig:doughnut_lens} but upon closer inspection they
can also be seen in several of the frames of
Fig.~\ref{brentjens_fig:rmcube} (e.g. RM= $-24$, 21 and 51).  Their
origin is still being investigated. They were present only in the
original Stokes $U$ (and $V$) images but the rotation of the
polarization vector, after de-applying the ionospheric Faraday
rotation, as well as the rotation inherent to the RM synthesis process
itself, leads to their appearance in every RM frame.


Finally, we also detect polarized grating lobes from the source
\object{Cas~A} in the input $Q$ and $U$ image cubes. However, due to
the great angular distance of this source from the field centre
($\approx$ 40$\degr$) its chromatic grating response moves rapidly
across the field of view as we move in frequency. In the final
RM-cube, these instrumental features therefore decrease to a level
well below the noise.


\subsection{Astronomical signals}
\label{brentjens_sec:astronomical_signals}

Once the instrumental features are recognized, the polarization images
and the RM-cube frames are seen to be filled with highly significant
signals of clearly celestial origin.  They cover a wide range of
Faraday depths $\phi$ between 0 and $90 \ \mbox{rad}\ \mbox{m}^{-2}$.
Broadly speaking there seem to be two types of emission:
\begin{enumerate}
\item diffuse structures with slowly changing polarization angles on
scales of the order of several tens of arc minutes: these have $0 \le
\phi \le 15\ \mbox{rad}\ \mbox{m}^{-2}$
\item distinct large structures having sizes of the order of a degree
and granularity in the polarization angle on scales of the order of a
few arc minutes: these have $30 \le \phi \le90 \ \mbox{rad} \
\mbox{m}^{-2}$
\end{enumerate}
There appears to be no significant emission between 15 and $30\
\mbox{rad}\ \mbox{m}^{-2}$.

\begin{figure*}
\centering
\includegraphics[width=\textwidth]{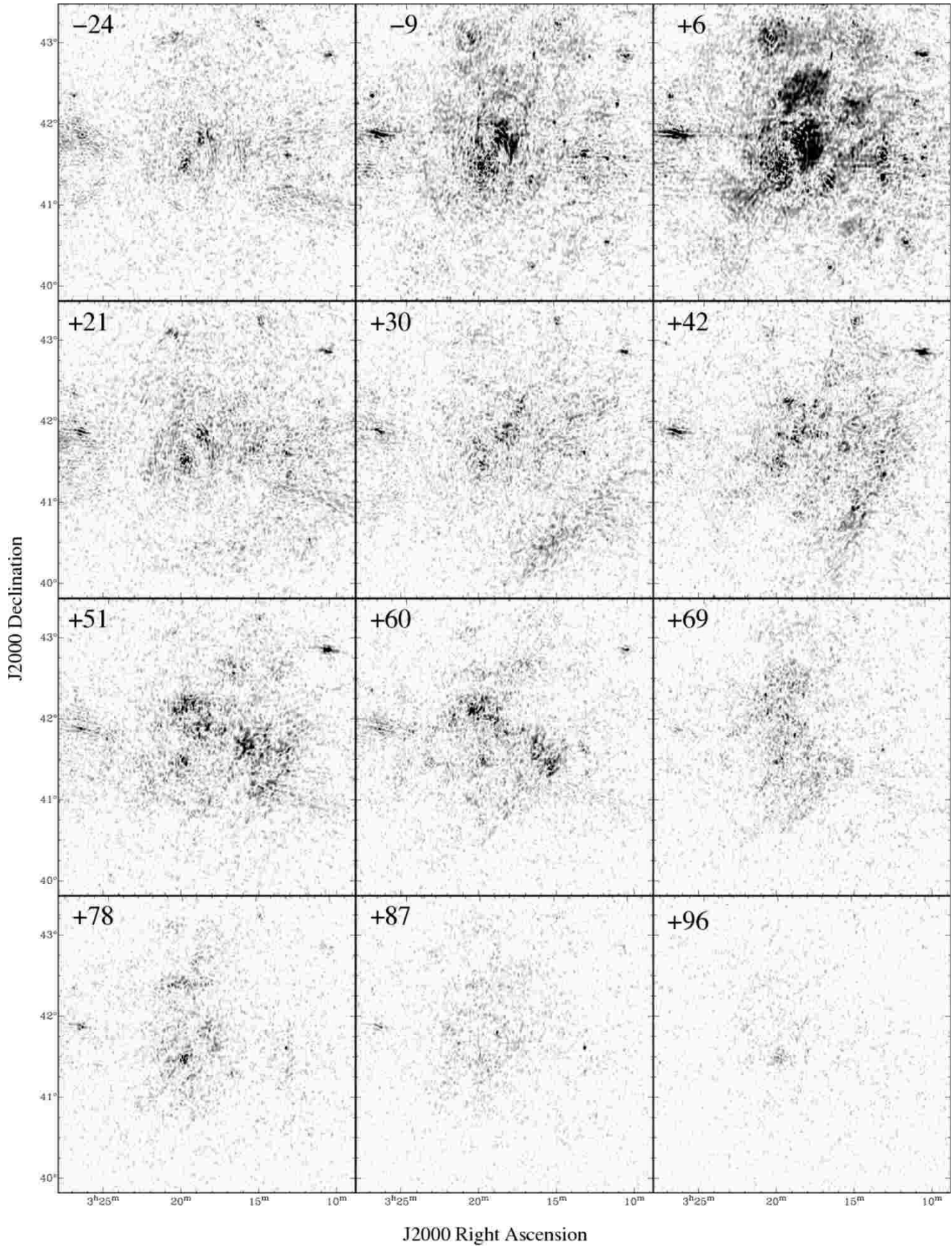}
\caption{Characteristic frames of polarized intensity ($\|P\|$) 
from the RM-cube. The Faraday depth of
each frame is specified in the top left corner in rad~m$^{-2}$.
}
\label{brentjens_fig:rmcube}
\end{figure*}

The first type of diffuse emission can be seen best in the third frame
of Fig.~\ref{brentjens_fig:rmcube}. This diffuse emission appears to
be spread over an area significantly wider than the primary beam
($2\fdg 2$--$2\fdg 6$) although it does decrease in brightness, as any
astronomical emission ought to. We emphasize this because the high
Faraday depth emission is notably more confined to the central area.
Fig.~\ref{brentjens_fig:foreground_Q} shows de-rotated Stokes $Q$ at a
Faraday depth of $+6$~rad~m$^{-2}$. It illustrates the large coherence
of the polarization angle across these patches.



\begin{figure}
\resizebox{\hsize}{!}{\includegraphics{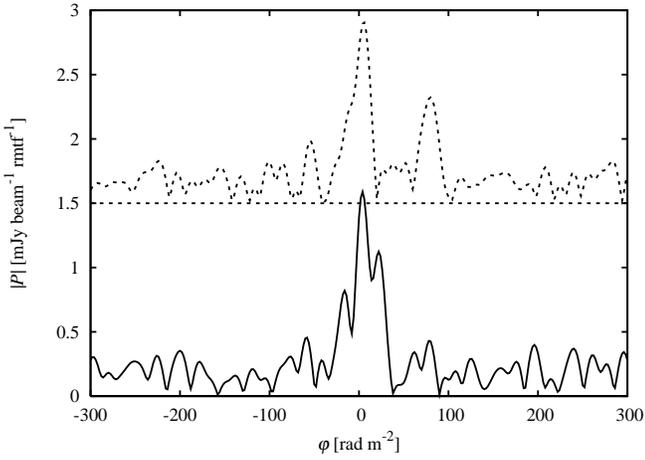}}
\caption{Faraday depth spectra of two regions of Galactic foreground
emission. The upper spectrum shows a region slightly west of the
bar. The lower spectrum shows a foreground patch near the lens.}
\label{brentjens_fig:foreground_spectrum}
\end{figure}


\section {Galactic foreground emission at low $\phi$}

\label{brentjens_sec:foreground} 

The large-scale diffuse polarized emission at low $\phi$ is very
similar to the features seen in previous WSRT studies at 92~cm.  This
component has been shown to be produced in the relatively local
Galactic medium, probably well within 1 kpc from the Sun
\citep{HaverkornEtAl2003b,Haverkorn2002,Wieringa1993}.  
This diffuse and complex polarization structure is believed to result
from the line-of-sight superposition of intrinsically highly polarized
emission and spatially as well as depth dependent Faraday rotation by
the interstellar (Galactic) magneto-ionic medium.
\citet{HaverkornKatgertDeBruyn2003} found that the rotation measure of
this Galactic foreground in Auriga ($l = 161\degr, b = +16\degr$) is
typically between $-17$ and $+10\ \mbox{rad}\ \mbox{m}^{-2}$, with an
average of $-3.4\ \mbox{rad}\ \mbox{m}^{-2}$. The values we find
towards the Perseus cluster at similar $l$ $(l\approx 150\degr)$ but
opposite $b$ $(b\approx -13\degr)$, are very similar. They cover the
range from 0 to 12$\ \mbox{rad}\ \mbox{m}^{-2}$.

Fig.~\ref{brentjens_fig:foreground_spectrum} displays a few typical
Faraday depth 'spectra' of Galactic foreground patches, in areas where
there is no significant emission due to (side lobes from)
instrumentally polarized emission.  It is also evident that at least
part of the foreground is `extended' or 'thick' in Faraday depth. Both
spectra suggest that part of the positive contribution to the Faraday
depth occurs inside the medium that produces the foreground emission.
One should nevertheless not be misguided by the height of the peaks in
a Faraday spectrum. The various artefacts in the individual $Q$ and
$U$ images that were described above occasionally conspire to produce
spurious peaks and one should therefore inspect the individual $Q$ and
$U$ images at the location of a peak in a RM-movie to verify the
reality of the RM-feature. Spatial continuity is a powerful aid in
ascertaining the reality of most features, and this is best done in
the RM-movie.



We find no exceptionally bright foreground features directly in front
of \object{3C~84} that could be associated with the low surface brightness
polarized emission that was detected in very high dynamic range 21~cm WSRT
observations \citep{Sijbring1993,DeBruynUnpublished1995}.
We will return to this in Section {\ref{brentjens_sec:thomson}}.


\section{Background structures at high $\phi$}
\label{brentjens_sec:background}

\subsection{Description of structure in the RM-cube}

\begin{figure}
\resizebox{\hsize}{!}{\includegraphics{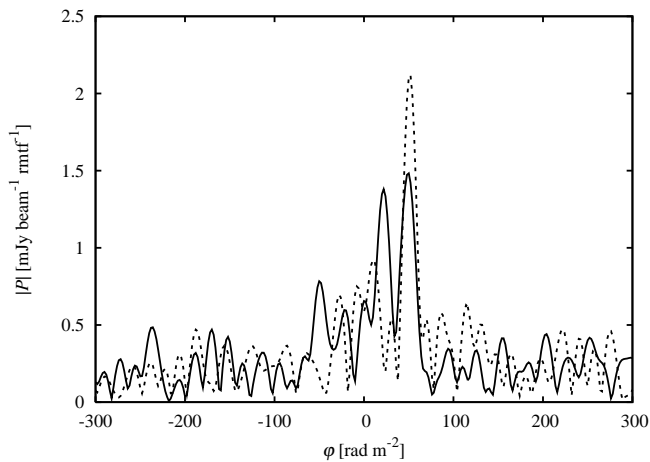}}
\caption{Faraday depth spectra of two lines-of-sight through the doughnut.}
\label{brentjens_fig:doughnut_spectrum}
\end{figure}

The second type of diffuse polarized emission is much richer in
spatial structures. Note that we will only discuss the polarized
intensity and defer to a future paper the discussion and analysis of
the rich structure in the polarization angle distribution.
The description starts with the fifth frame of Fig.~\ref{brentjens_fig:rmcube}
($\phi = 30 \ \mbox{rad}\ \mbox{m}^{-2}$) which shows a weak, front-like
structure running from $\alpha \approx 3^\mathrm{h}16^\mathrm{m},
\delta \approx 40\degr 24\arcmin$ to $\alpha \approx
3^\mathrm{h}10^\mathrm{m}, \delta \approx 41\degr 36\arcmin$. The next
frame ($\phi = 42\ \mbox{rad}\ \mbox{m}^{-2}$) displays a stronger
linear feature with a slightly different position angle. A bright
circular 'doughnut shaped' structure, with a diameter of about
7$\arcmin$ develops at position $\alpha \approx 3^\mathrm{h}
15^\mathrm{m} 35^\mathrm{s}, \delta \approx 41\degr 42\farcm
3$. Fig.~\ref{brentjens_fig:doughnut_spectrum} shows Faraday depth
spectra of two lines-of-sight through the doughnut. The dashed
spectrum shows the high significance of the doughnut. The peak at
negative Faraday depth in the solid spectrum is caused by a
whisker. The bright peak near $\phi=+10$~rad~m$^{-2}$ is due to the
Galactic foreground.

\begin{figure}
\resizebox{\hsize}{!}{\includegraphics{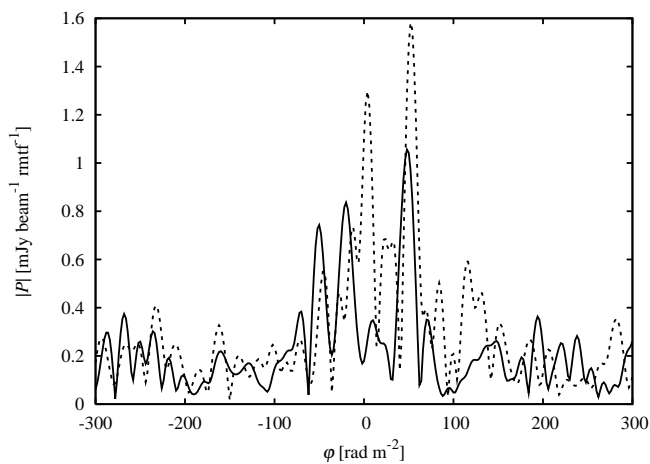}}
\caption{Faraday depth spectra of two lines-of-sight through the lens.}
\label{brentjens_fig:lens_spectrum}
\end{figure}

 Frame number seven ($\phi = 51 \ \mbox{rad} \
\mbox{m}^{-2}$) shows a spectacular lenticular feature southwest of
the doughnut. It is even better seen in
Fig.~\ref{brentjens_fig:doughnut_lens}, which is an enlarged frame
from the RM-cube at $\phi=+52$~rad~m$^{-2}$. The position angle of the
lens is very similar to the position angle of the linear structure in
the previous frame. The lens is roughly $40\arcmin\times 20\arcmin$ in
size. At the distance of the Perseus cluster, it translates to
$1\times 0.5 \ \mbox{Mpc}$. Two Faraday depth spectra through the lens
are shown in Fig.~\ref{brentjens_fig:lens_spectrum}. The two peaks at
negative $\phi$ in the solid spectrum are caused by instrumental
whiskers. The emission from the lens peaks at $\phi\approx
52$~rad~m$^{-2}$. This frame and the next one, frame 8 at $\phi$ = 60
rad m$^{-2}$, also show bright extended emission with a polarized
intensity of about 1 mJy~beam$^{-1}$ approximately $40\arcmin$ north
of \object{3C~84}. This patch is roughly $15\arcmin \times 25\arcmin$
and is located north-west of the steep spectrum tail of
\object{NGC~1265} \citep{SijbringDeBruyn1998}. 

\begin{figure}
\resizebox{\hsize}{!}{\includegraphics{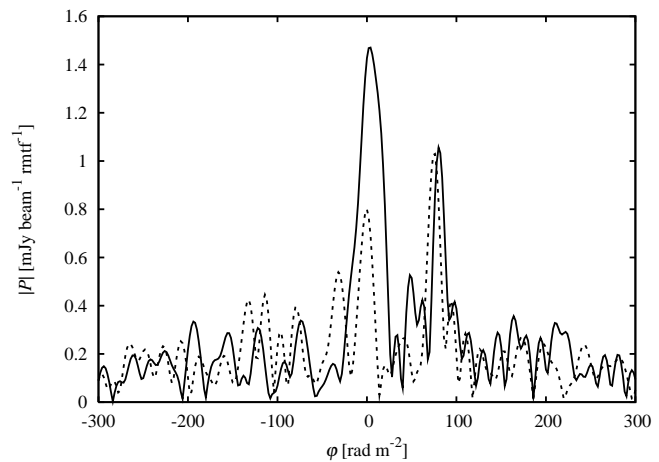}}
\caption{Faraday depth spectra of two lines-of-sight through the straight bar
  north east of \object{NGC~1265}.}
\label{brentjens_fig:bar_spectrum}
\end{figure}

The lens and doughnut slowly fade away in the frames that follow and
polarized emission disappears from the western part of the
field. Frame nine shows mottled emission across an area of
$2\degr\times 1\fdg5$ centred around the area between \object{NGC
1275} (\object{3C~84}) and \object{NGC~1265}.  The tenth frame ($\phi =
78\ \mbox{rad}\ \mbox{m}^{-2}$), is the last frame that shows
significant emission. The horizontal bar almost $1\degr$ due north of
\object{3C~84}, at $\alpha \approx 3^\mathrm{h}20^\mathrm{m}, \delta
\approx 42\degr 25\arcmin$, is very conspicuous at these values of
$\phi$. Two Faraday depth spectra are shown in
Fig.~\ref{brentjens_fig:bar_spectrum}. The emission fades away towards
the cluster centre. Beyond $\phi \approx 100\ \mbox{rad}\
\mbox{m}^{-2}$ no structures are detectable at $2\arcmin$--$3\arcmin$
resolution.


\subsection{Where is the total intensity counterpart !?}
\label{brentjens_sec:where_is_total_intensity}

We have not yet detected any of the extended polarized features in our
RM-cube in the total intensity image.  This was rather surprising
because we have no doubt that this emission is due to the synchrotron
process. There may, however, be a rather simple explanation.  The
sensitivity in the Stokes $I$ image is about a factor 10--15 poorer
than in the polarization image: 1.5~mJy~beam$^{-1}$ as against 0.1
mJy~beam$^{-1}$ (both numbers now refer to the $2\arcmin\times
3\arcmin$ resolution image. However, the brightest features in the
polarization images, e.g. the `lens' and the 'doughnut', have a peak
brightness of about 1~mJy~beam$^{-1}$. So the lack of detection in $I$
implies, at face value, a polarization percentage of at least
50\%. Although this is not impossible for synchrotron emission, these
percentages are getting 'uncomfortably' high. Several 'relic'
structures in clusters of galaxies \citep{EnsslinEtAl1998} have
polarization percentages of 20--30\% although these all refer to
shorter wavelengths where depolarization is less of an issue.


Apparent polarization percentages well above 100\% are quite normal in
the case of Galactic foreground polarization \citep{Wieringa1993,
HaverkornEtAl2003b,Haverkorn2002,GaenslerEtAl2001}.  They are due to
the fact that the Galactic foreground Stokes I is very smooth and
largely resolved out due to the lack of sufficiently short
interferometer spacings. The emission in Stokes $Q$ and $U$, on the
other hand, has high spatial frequencies because they can be both
positive and negative (cf. Fig.~\ref{brentjens_fig:foreground_Q}).
Fig.~\ref{brentjens_fig:rm_maxmap}, to which we will return later,
shows that the distribution of polarized intensity indeed has a
significant large-scale component. If the total and polarized
intensity would be distributed similarly, the shortest spacings of the
WSRT will have only partly sampled the signals. This was exacerbated
by the fact that in two out of six usable frequency bands one of the
crucial telescopes (RT9 and RTA) providing the shortest 36~m spacing
were not functioning properly.  The true polarization percentage of
the polarization structures could then easily be a factor 2--3 lower
which would decrease the apparent percentage polarization from $>$50\%
to $>$20\%. The conclusion, however, remains that the signals must be
highly polarized. We will return to this aspect below.



\section{Discussion}
\label{brentjens_sec:discussion}

\subsection{Location of high Faraday depth features}

There are two possible locations for the source of the emission at
$\phi \ge 30 \ \mbox{rad} \ \mbox{m}^{-2}$. The first possibility is
that here too we are dealing with Galactic synchrotron emission.
Potential sites are the region between us, and possibly including, the
Perseus arm which is at a distance of 2--3 kpc. The second possible
location is the Perseus cluster of galaxies, which was the central
target of these observations.  We favour the second option for a
number of reasons:
\begin{itemize}
\item the high $\phi$ structures seem to be largely confined to and
approximately centred at the Perseus cluster, contrary to the low
$\phi$ emission;
\item  there is a systematic decrease in the small scale structure in
Stokes $Q$ and $U$ as a function of distance to the cluster centre.
\end{itemize}
Initially we believed there was a third argument for an association
with the Perseus cluster: when scanning the RM-movie there is a very
suggestive systematic convergence of the higher RM values towards the
centre of the Perseus cluster. However, we could not exclude that this
was caused in part by an anomaly in the Galactic foreground RM
distribution.

\citet{JohnstonHollittHollittEkers2004} have constructed an interpolated
map of the RM distribution of a sample of over 800 radio sources, from
which we may estimate the Galactic RM contribution in any
direction. The expected RM near the \object{Perseus cluster} is
somewhere between $+10$ and $+20\ \mbox{rad}\ \mbox{m}^{-2}$. The RM
expected from the large-scale Galactic field should be negative in
this area. The positive values inferred by
\citet{JohnstonHollittHollittEkers2004} may therefore indicate an RM
anomaly south of the Galactic Perseus arm. 


\subsection{Evidence for a Galactic RM gradient}

\begin{figure}
\centering
\includegraphics[width=\hsize]{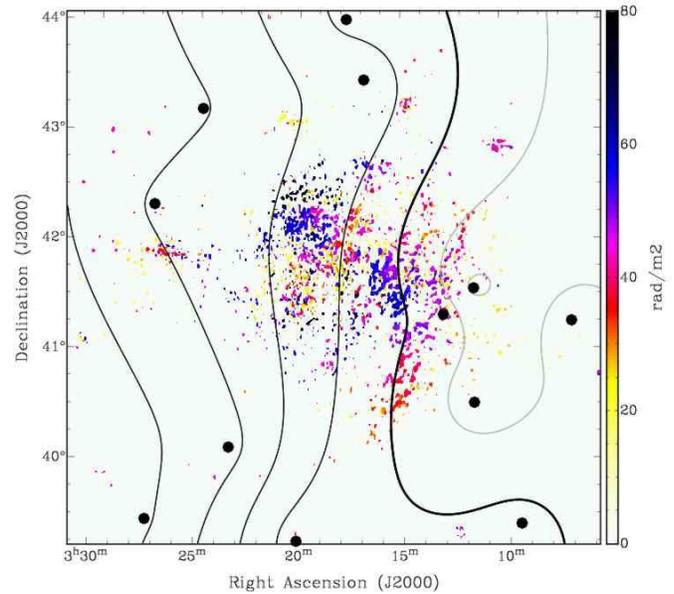}
\caption{Faraday depth of the maximum $|P|$ along the line-of-sight,
excluding range 3--19 inclusive. Pixels with a peak polarized flux
density less than 0.7 mJy per beam have been masked out.
Overplotted is a radial basis function
interpolation \citep{CarrEtAl2001} of the ``foreground RM'' determined
using the 12 sources indicated with filled circles. The thick line has
a RM = 0 rad m$^{-2}$, and contours are given in steps of 20 rad m$^{-2}$ with
positive values increasing to the east.} 
\label{brentjens_fig:rm_maxmap}
\end{figure}

To investigate this further and put our suspicions about foreground
contamination to the test we proposed a new series of observations in
August 2004, when most of this work had already been finished and
written up. Using the WSRT at 21~cm we conducted a series of snapshot
observations of 15 polarized (background) radio sources in the
direction of the Perseus cluster. The sources were selected through
the NVSS \citep{CondonEtAl1998} and should have at least a few mJy of
polarized emission.  Using the 160~MHz wide WSRT band this appeared to
be sufficient to allow a RM determination accurate to better than a
few rad m$^{-2}$ for all 15 sources after 30~m of integration per
source.  After excluding three sources with a complicated brightness
morphology and multiple, widely different RMs, the remaining 12 sources
showed a smooth gradient across the field of view. The results are
shown in Fig.~\ref{brentjens_fig:rm_maxmap} and include a radial
basis-function interpolation of a smooth foreground screen
\citep{CarrEtAl2001}.  Because the uncertainty in the RM determination
of each source is at most a few rad m$^{-2}$, the gradient is very
significant.  We also infer that the scatter due to a RM contribution
from the medium inside, or surrounding, the radio sources can be at
most 5--10 rad m$^{-2}$; this is consistent with previous studies of
RM differences in double radio sources
\citep{SimonettiEtAl1984,SimonettiEtAl1986,Leahy1987}.

This smooth Galactic foreground RM gradient should be compared with the 
observed spatial distribution in the diffuse high Faraday depth 
emission features. To do this we proceeded as follows but we begin
with noting that there is no such thing as \emph{the} RM or \emph{the}
Faraday depth of a pixel: indeed there can be multiple regions of
emission at different Faraday depths along the line-of-sight
(for an extensive discussion of such situations, which are quite 
normal at low frequencies, we refer to 
\citet{BrentjensDeBruyn2005}). Within the RM-cube we masked out the
emission at $3 \le \phi \le 19\ \mbox{rad}\ \mbox{m}^{-2}$ which is
very clearly dominated by the Galactic foreground.  At every pixel in
the RM-cube we then determined at which Faraday depth the polarized
intensity $|P|$ was maximal.  This range included $\phi = 0\
\mbox{rad}\ \mbox{m}^{-2}$, the value where all frequency independent
instrumental problems accumulate. This guaranteed that the emission
that would be assigned a Faraday depth other than zero, cannot be caused by
RMTF side lobes of instrumental problems. The regions where the
latter is manifestly the case are the sources at the edges of the 
field. The result of this exercise is shown in 
Fig.~\ref{brentjens_fig:rm_maxmap}\footnote{The on-line version of
this article contains a colour version of this figure.}.
We imposed a cut at $|P| < 0.7\ \mbox{mJy}\ \mbox{beam}^{-1}$ before
including pixels.


Fig.~\ref{brentjens_fig:rm_maxmap} reveals that across the 2$\degr$
diameter area with high $\phi$ emission there is a clear Galactic RM
gradient of about 60~rad~m$^{-2}$, which is approximately equal to the
observed range of $\phi$ = 30--90~rad~m$^{-2}$. In other words, the
high $\phi$ emission shows a fairly uniform positive excess of about
40--50 rad m$^{-2}$ relative to the interpolated Galactic RM.  For all
but two of the background sources the line-of-sight avoids the Perseus
cluster, which we define as the area out to which significant X-ray
emission can be seen, a radius of about 1$\degr$. We may therefore
assume that the RM of these background radio sources represents the
line-of sight-integrated RM of our Galactic foreground magneto-ionic
medium.  This result holds the key to the interpretation of the high
$\phi$ emission as we will now proceed to argue.


For the sake of the argument let us concentrate on the features on the
western side of the cluster near what we have called the shock, the
lens and the doughnut.  The diffuse organized Galactic polarized
emission has a RM of about +10 rad m$^{-2}$. In order for the total,
integrated, RM to the background sources to end up at about 0 rad
m$^{-2}$ there must be another small negative contribution of about
$-10$~rad~m$^{-2}$ from somewhere along the line-of-sight in our
Galaxy.  This seems not implausible. The RM-cube around $\phi \approx
0$~rad~m$^{-2}$ is rather messy due to the remaining instrumental
polarization from the extremely bright and complex cluster radio
sources.  Now let us suppose that the 'screen' responsible for the
40--50~rad~m$^{-2}$ contribution is also due to our Galaxy and extends
across the full field of view shown in
Fig.~\ref{brentjens_fig:rm_maxmap} (the sudden termination at a
distance of about $1\fdg 5$ from the pointing centre must then be
attributed largely to the primary beam attenuation, which is unlikely
but not impossible).  We would then require another screen, but now
with a Faraday depth of $\phi$ = $-40$~rad~m$^{-2}$ to compensate for
the $\phi$ = $+40$~rad~m$^{-2}$ screen.  We see no evidence in our
RM-cube for any emission from such a screen.

Even stronger arguments can be brought forward for the case that the
$\phi$ = +40 rad m$^{-2}$ emission would result from a discrete cloud
in the Perseus arm of our Galaxy that just happened to cover the
Perseus cluster. Assuming a magnetic field of 1~$\mu$Gauss, a depth of
about 60 parsec (equal to a lateral scale of $1\fdg 5$ at 2.5 kpc), a
RM of 40~rad~m$^{-2}$ requires a (smooth) electron density of
0.8~cm$^{-3}$.  Assuming roughly 20\% H$\alpha$ extinction, this
density and path length would generate an $\mathrm{H}\alpha$ surface
brightness of about 13 Rayleighs. The total H$\alpha$ surface
brightness towards the Perseus cluster in the WHAM survey
\citep{HaffnerEtAl2003} is about 4 Rayleighs. Furthermore, the integrated
$\mathrm{H}\alpha$ emission is rather smooth towards the Perseus
cluster, ranging in surface brightness from 3--5 Rayleighs over an
area of $3\degr\times3\degr$centred on the Perseus cluster. There is
no excess at the location of the cluster. The discrete Perseus arm
model for the origin of the excess $+40$~rad~m$^{-2}$ polarized
emission therefore appears to be untenable as well.

A similar argument can be given for the emission to the north
and east of the cluster centre. The emission in this
region shows a significant drop in surface brightness well before 
the primary beam attenuation sets in. This would have to be a
peculiar coincidence in the case of a Galactic foreground origin. 

When we add the above arguments to those given previously, the spatial
coincidence with the cluster and the granularity of the polarization
angle structure, we are led to the conclusion that the high $\phi$ 
emission indeed must be associated with the Perseus cluster of galaxies.

\subsection{Intracluster or peripheral emission?}
\label{brentjens_sec:peripheral_emission}

Having argued that the high $\phi$ emission is associated with the
Perseus cluster we may then ask the question: \emph{where} in the
cluster does the emission originate?  Does it originate from within
the cluster or is it emitted from the periphery of the cluster? 
At this point we should also realize that this question really has two
aspects: 
\begin{enumerate}
\item Where does the emission come from?
\item Where does the Faraday rotation occur?
\end{enumerate}

\begin{figure}
\resizebox{\hsize}{!}{\includegraphics{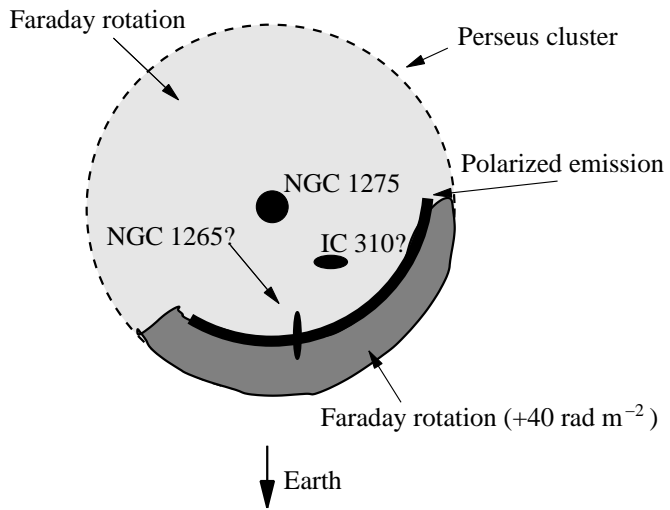}}
\caption{Sketch of our proposal for the situation in the Perseus
cluster. A top view of the distribution in the Perseus cluster of
various emitting areas and Faraday rotating areas.}
\label{brentjens_fig:cartoon_perseus}
\end{figure}

Fig.~\ref{brentjens_fig:cartoon_perseus} is a sketch that is helpful
for the following discussion. It shows where we believe the emission
and rotation occur in the Perseus cluster. From the fact that most of
the polarized structures are unresolved by the RMTF it appears that
the medium is not Faraday-thick (see
\citet{BrentjensDeBruyn2005}). This means that is the largest fraction
of the Faraday rotation of 40 rad m$^{-2}$ occurs between the emitting
region and us.

At this point it is relevant to recall that there are two extended and
polarized radio sources within the Perseus cluster, \object{NGC~1265}
and \object{IC~310}, in the area where we find diffuse polarized
emission.  Their head-tail morphology indicates that they are most
likely located within the denser gaseous part of the Perseus cluster
\citep{MileyEtAl1972}.  The RM across the bright part of the tail of
\object{NGC~1265} \citep{ODeaOwen1986, ODeaOwen1987} shows a scatter
of 20~rad~m$^{-2}$ around a mean value of 45--50~rad~m$^{-2}$. The
scatter is probably due to the magnetized plasma in the cocoon
surrounding and mixed within the twin tails of this head-tail source.
The 'expected' value due to our Galactic foreground at the location of
\object{NGC~1265} is about 25~rad~m$^{-2}$
(cf. Fig.~\ref{brentjens_fig:rm_maxmap}).  The difference of about 20
rad m$^{-2}$ must then be attributed to the medium in the
line-of-sight between the head-tail source and the edge of our Galaxy:
i.e. the peripheral region of the near side of the cluster.

\begin{figure}
\resizebox{\hsize}{!}{\includegraphics{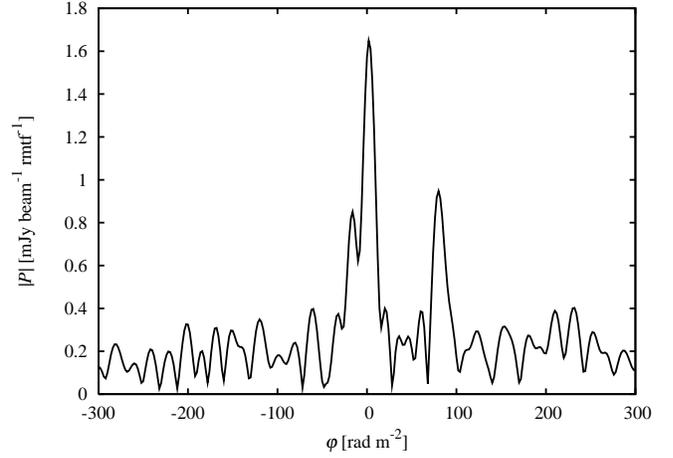}}
\caption{Faraday depth spectrum near the 'head' of \object{IC~310}. A highly
significant feature at Faraday depth $\phi$ = 80~rad~m$^{-2}$ can be
detected. The peak around $\phi$ = 0~rad~m$^{-2}$ is due to the
instrumental polarization of \object{IC~310}, which has a peak in
total intensity of more than 500 mJy.}
\label{brentjens_fig:IC310.spectrum}
\end{figure}

We also detect faint but highly significant polarized emission from a
region close to the 'head' of \object{IC~310} at a value of $\phi$ =
+80~rad~m$^{-2}$ (Fig.~\ref{brentjens_fig:IC310.spectrum}). This
polarized emission probably originates from a region about 1$\arcmin$
downstream from the head of \object{IC~310} where significant
polarized emission was also detected at 610~MHz by
\citet{SijbringDeBruyn1998}. This is sufficiently far away from the
galaxy that we may assume that the Faraday rotation is not due to the
ISM associated with \object{IC~310} itself. The expected Galactic RM
value (cf. Fig.~\ref{brentjens_fig:rm_maxmap}) at this location is
about +20~rad~m$^{-2}$.  The difference of +60~rad~m$^{-2}$ could
imply that
\object{IC~310} is located deep within the cluster.  

Note that the excess Faraday depth shown by \object{NGC~1265} and
\object{IC~310} is consistent with the average cluster RM excess of
less than 50 rad m$^{-2}$ found by \citet{ClarkeEtAl2001} at the
projected distance of 700 kpc of these two head-tail radio sources
from the centre of the Perseus cluster.
The polarized emission having $\phi$= 40 rad m$^{-2}$ is  
probably located on the periphery on the near side of the cluster
at about the same depth as  \object{NGC~1265}. \object{IC~310}
may well be located \emph{behind} the diffuse polarized emission.


\subsection{Thomson scattering within the Perseus cluster?}

\label{brentjens_sec:thomson}


\begin{figure}
\resizebox{\hsize}{!}{\includegraphics{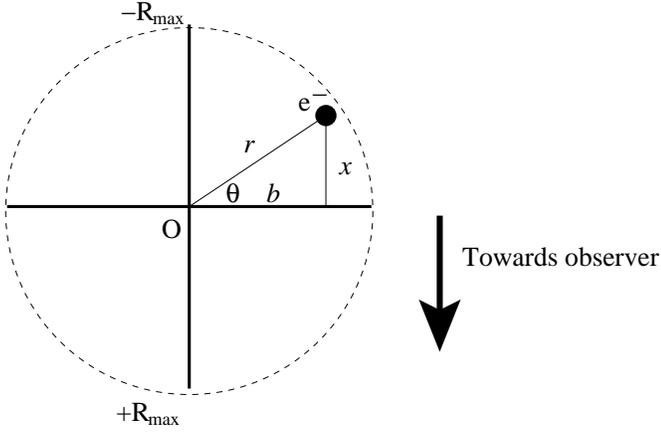}}
\caption{Thomson scattering geometry.}
\label{brentjens_fig:thomson_geometry}
\end{figure}

As discussed in the introduction the initial goal of the observations
was to characterize the possible contamination of Galactic foreground
polarization to the observed 21~cm polarization
\citep{DeBruynUnpublished1995} which was observed to straddle 
\object{3C~84} on an
angular scale of 30$\arcmin$ and believed to be due to Thomson
scattering. In a future paper we plan to present the original 21~cm
observations augmented with a wider field study.

Could the polarized emission that we detected at wavelengths of 
81--95~cm be indeed Thomson scattered radiation from \object{3C~84} 
or its direct surroundings
\citep{Syunyaev1982, WiseSarazin1990,WiseSarazin1992}\footnote{Please
note that \citet{WiseSarazin1990} made a small error in their equation
(1), which makes it incompatible with our equations
(\ref{brentjens_eqn:thomson_xb_integral}) and
(\ref{brentjens_eqn:thomson_knu}). Equation (1) of
\citet{WiseSarazin1990} must be multiplied by $1/4\pi$ in
order to correct this. The same error is also present in equation
(2.1) of \citet{WiseSarazin1992}, hence their estimates are one order
of magnitude too optimistic.}?

In order to quantitatively answer that question, we have made a simple
model of the Perseus cluster. The redshift of the Perseus cluster is $z=0.0167$
\citep{StrubleRood1999}. We assume $H_0 = 72\pm2\ \mbox{km}\
\mbox{s}^{-1}\ \mbox{Mpc}^{-1}$ \citep{SpergelEtAl2003} which gives 
a distance to the cluster of $69.5\pm 2$ Mpc. We have adopted a spherically
symmetric electron distribution. We used the deprojected radial
dependence determined by \citet{ChurazovEtAl2003}:

\begin{equation}
n_\mathrm{e} = {{3.9\times10^{-2}}\over{\left[1+\left({r \over
80\ \mathrm{kpc}}\right)^2\right]^{1.8}}} +
{{4.05\times10^{-3}}\over{\left[1+\left({r\over
280\ \mathrm{kpc}}\right)^2\right]^{0.87}}}\ \mbox{cm}^{-3}
\label{brentjens_eqn:ne_perseus_churazov}
\end{equation}

In order to estimate the radio radiation, we summed the contributions
of the \object{3C~84} $30\arcsec$ component and the halo. We ignore
the very bright core, because it has an uncertain history and its high
flux density is largely due to an outburst that only started in the
late 1950's \citep{ODea1984}. Its Thomson echo cannot therefore have
spread by more than a few tens of parsecs.  The values for the
luminosity from the extended components are taken from
\citet{Sijbring1993} after scaling them to our adopted distance. Both
source components were situated in the centre of gravity of the
electron population. We modelled the source as a single unpolarized
point source and assumed an isotropic radiation field.

%

\begin{eqnarray}
P(\nu') & = &
10.2\times10^{24}\left({{\nu'}\over{333\ \mathrm{MHz}}}\right)^{-1.15}
+\\
\nonumber & &
9.83\times10^{24}\left({{\nu'}\over{333\ \mathrm{MHz}}}\right)^{-1.2} 
\ \mathrm{W}\ \mbox{Hz}^{-1},
\end{eqnarray}
where $\nu' = (1+z)\nu$ is the rest frame frequency of the Perseus cluster.


\begin{figure}
\resizebox{\hsize}{!}{\includegraphics{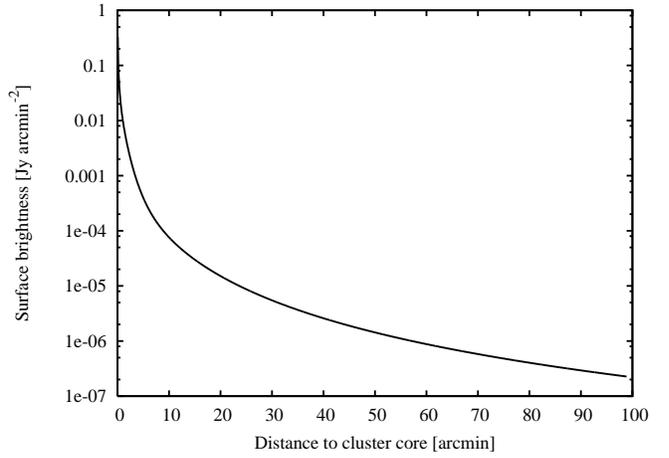}}
\caption{Computed Thomson scattering halo surface brightness at 92~cm 
based on the radio flux density of the \object{3C~84} halo 
and $30\arcsec$ component. Total intensity.}
\label{brentjens_fig:thomson_halo_IP}
\end{figure}


\begin{figure}
\resizebox{\hsize}{!}{\includegraphics{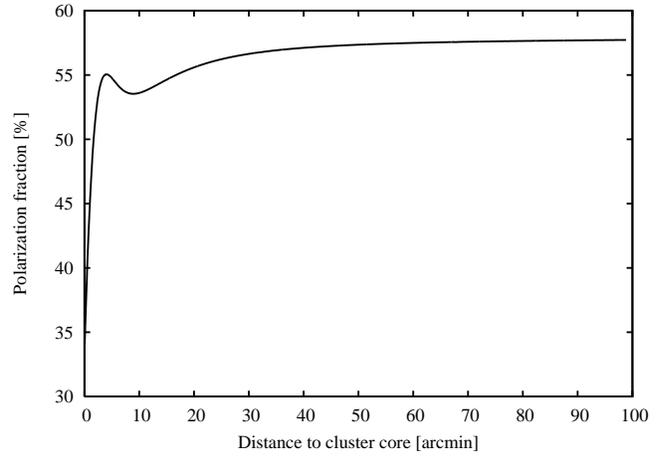}}
\caption{Computed Thomson scattering halo surface brightness 
at 92~cm based on the radio flux density of the \object{3C~84} halo 
and $30\arcsec$ component. Polarization fraction.}
\label{brentjens_fig:thomson_polfraction}
\end{figure}

The geometry of the computation is given in
Fig.~\ref{brentjens_fig:thomson_geometry}. The radio source is located
at the origin ($O$). The emission is scattered at location ($b,
x$). The total intensity of the scattered emission is given by the
line integral

\begin{equation}
I(\nu) = \int_{-\infty}^{+\infty} \frac{P((1+z)\nu)}{4\pi
r^2}\frac{\mathrm{d}\sigma}{\mathrm{d}\Omega}
n_\mathrm{e}(\mathbf{r}(x,b))\mathrm{d}x,
\label{brentjens_eqn:thomson_scattering_general}
\end{equation}

\noindent where

\begin{equation}
\frac{\mathrm{d}\sigma}{\mathrm{d}\Omega} = \frac{1}{2}r_0^2\left(1+\sin^2\theta\right)
\end{equation}

\noindent and

\begin{equation}
r_0^2 = \frac{3}{8\pi}\sigma_\mathrm{T}.
\end{equation}

\citep{Syunyaev1982, RybickyLightman}. We can write
equation (\ref{brentjens_eqn:thomson_scattering_general}) in terms of $b$
and $x$ as depicted in Fig.~\ref{brentjens_fig:thomson_geometry}.

\begin{equation}
I(\nu) = K(\nu)
\int_{-L}^{+L} n_\mathrm{e}(r){{b^2 + 2x^2}\over{\left(b^2 +
x^2\right)^2}}\mathrm{d}x
\ \mathrm{W}\ \mbox{Hz}^{-1}\ \mbox{m}^{-2}\ \mathrm{Sr}^{-1},
\label{brentjens_eqn:thomson_xb_integral}
\end{equation}
where $b$ is the impact parameter, $x$ is the path length measured from
the midplane of the electron distribution, $r = \sqrt{b^2+x^2}$ is the
distance from the radio source, $L = \sqrt{R^2_\mathrm{max} - b^2}$ is
the integration boundary, and

\begin{equation}
K(\nu) = {{3P((1+z)\nu)\sigma_\mathrm{T}}\over{64\pi^2(1+z)^3}} 
\ \mbox{W}\ \mbox{m}^{2}\ \mbox{Hz}^{-1}\ \mbox{Sr}^{-1},
\label{brentjens_eqn:thomson_knu}
\end{equation}
independent of position. We used SI units throughout this computation.
Assuming there is no Faraday rotation in the cluster medium, which of
course is unrealistic at a wavelength of 92~cm, we may also compute
the maximum degree of polarization by

\begin{equation}
I_\mathrm{P}(\nu) = K(\nu)
\int_{-L}^{+L} n_\mathrm{e}(r){{b^2}\over{\left(b^2 +
x^2\right)^2}}\mathrm{d}x
\ \mbox{W}\ \mbox{Hz}^{-1}\ \mbox{m}^{-2}\ \mbox{Sr}^{-1}.
\end{equation}

The results are shown in Fig.~\ref{brentjens_fig:thomson_halo_IP} and
Fig.~\ref{brentjens_fig:thomson_polfraction}.
Fig.~\ref{brentjens_fig:thomson_halo_IP} indicates that at small
projected distances ($<30\arcmin$) from the cluster core it is, in
principle, possible to detect the Thomson scattering emission even at
92~cm wavelength. For example, at $25\arcmin$ distance from the cluster
centre, the expected total intensity is approximately
50~$\mu$Jy~$2\arcmin\times 3\arcmin$beam$^{-1}$. Assuming 50\%
polarization, this amounts to a polarized intensity of
25~$\mu$Jy~beam$^{-1}$. Barring dynamic range issues, it will however
be very difficult to separate this scattered emission from true
cluster halo emission in Stokes $I$. The much deeper sensitivity
reached in polarized intensity (70~$\mu$Jy~beam$^{-1}$) affords a
better chance if we can correctly estimate the complications from
Faraday rotation within the scattering medium.  So although we can not
exclude that part of the mottled polarized emission close to
\object{3C~84}, at the location where we previously observed
significant polarized emission at 21~cm, is due to Thomson scattering,
we will not discuss it here any further and will concentrate on the
brighter polarized features.

We believe that neither the large structures that we detected in the
western part of the field, nor the bar feature to the north, can be
explained by Thomson scattering.  The high percentage of polarization
inferred for the diffuse structures are consistent with Thomson
scattering (without significant de-polarization), but they appear far
too bright for their projected distance from \object{3C~84}.  For
example, to explain the observed polarized surface brightness of the
lens (approximately 1 mJy per \mbox{2$\arcmin\times3\arcmin$ beam}) by
Thomson scattering, either the source luminosity 3 to 6 million years
ago should have been a factor of 100 to 500 higher, or the electron
density at 1 to 2 Mpc distance from the cluster centre should be at
least $5\times10^{-3}\
\mbox{cm}^{-3}$. Both options, or a combination of a higher luminosity
and a higher density, appear to be implausible.  However, we note that
the current core luminosity is 5$\times$ higher than the luminosity
assumed in the computations and the uncertainty about the long term
average of the core emission will remain one of the uncertain issues
with the Thomson scattering interpretation.  In a future paper we will
return to the issue of the separation of Thomson scattered emission
from other Perseus cluster contributions using both existing and new
21~cm, 92~cm, and 200~cm WSRT observations.


\subsection{Structure formation shocks and AGN bubbles}
\label{brentjens_sec:agn}


\begin{figure}
\resizebox{\hsize}{!}{\includegraphics{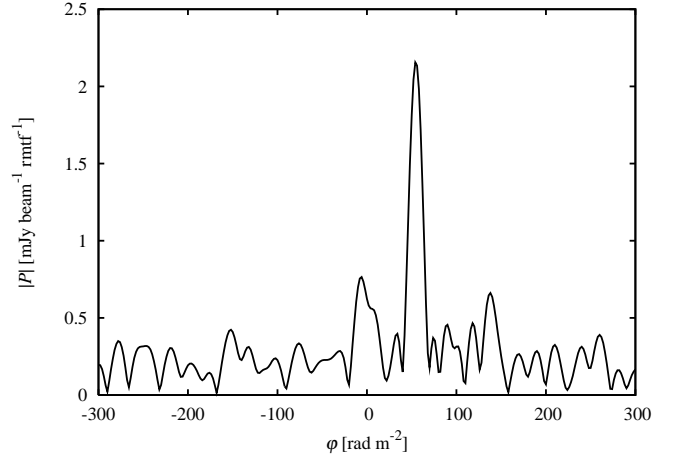}}
\caption{Faraday depth spectrum a line-of-sight through the curved
blob due north of \object{NGC~1265}.}
\label{brentjens_fig:blob1265}
\end{figure}

We will now briefly discuss the implications of what we consider to be
the most likely explanation for the nature of the diffuse high Faraday
depth structures centred on the Perseus cluster. A more in-depth study
will be presented in future papers.  We will follow
\citet{EnsslinEtAl1998} and \citet{EnsslinGopalKrishna2001} and
interpret the emission as resulting from (re-energized) relics and
emission associated with shocks in the large-scale structure formation
gas flow.

Cosmological simulations of structure formation have improved
dramatically in recent years. Nearly all simulations indicate
that structure in the Universe formed hierarchically. Small clumps
collapsed, larger ones started accreting smaller ones and grew even
larger. A consequence of this type of structure formation is the
presence of large-scale gas flows, which form huge, multi megaparsec
scale shocks at their intersections \citep{Burns1998,
QuilisIbanezSaez1998, Miniati2000}. 
\citet{EnsslinEtAl1998} discuss the possibility that cluster relic
sources, large highly polarized radio sources in the outskirts of
galaxy clusters, could trace interface shocks between clusters and the
super cluster filaments accreting onto them. The electrons in the relic
would be re-energized by diffusive shock
acceleration. \citet{EnsslinGopalKrishna2001} developed an analytic
formalism for an alternative process. They propose that cluster relics
are bubbles of magnetized plasma, released into the IGM by AGN. The
electrons in such a bubble loose energy through adiabatic expansion of
the bubble. Once fully detached, the bubble enters a 'hibernation'
phase in which it can stay for several Gyr; the only energy losses
suffered by the relativistic electrons are those due to Inverse
Compton losses off the 2.7 K background radiation. The bubble is now
buoyant in the IGM, due to its lower density \citep{Brueggen2003}.  It
is finally revived when the bubble is adiabatically compressed by a
large-scale structure formation shock
\citep{EnsslinBrueggen2002}.

The morphology of the long linear structures that appear in frames 5
and 6 of Fig.~\ref{brentjens_fig:rmcube} are indeed reminiscent of
shock fronts. This structure would then have to be located at the
periphery of the cluster. As argued in section
\ref{brentjens_sec:peripheral_emission} it would probably have to be
located on the near side of the cluster to prevent significant
beam depolarization following the passage of the radiation through the
Perseus cluster.  The good alignment of the front (frame 6) and the
lenticular structure (frame 7) argues that they are co-located. Do we
perhaps observe the actual compression of a huge ($1\times 0.5\
\mbox{Mpc}$) hibernating bubble by a large-scale structure formation
shock?

The \object{Perseus cluster} is situated at the eastern end point of
the Perseus-Pisces super cluster. The fronts and the lens are in the
area where one would expect large-scale structure formation shocks
from general structure formation simulations.  The fact that the front
is so straight over a distance of more than 3 Mpc is not impossible;
many of the fronts in these simulations show that behaviour.
\citet{EnsslinEtAl2001} report a tentative detection of a cosmological
shock wave that overruns one of the radio lobes of
\object{NGC~315}. \object{NGC~315} is positioned at the intersection
of several sub filaments of the Perseus-Pisces super cluster.

The structure of the doughnut strongly resembles the structure and
topology of the magnetic field in a pre-shock bubble, as
simulated by \citet{EnsslinBrueggen2002}. Assuming the doughnut to be
situated at the distance of the Perseus cluster, it measures 250
kpc across, consistent with the hypothesis of an AGN bubble.

The blob that is visible in frame 7 of
Fig.~\ref{brentjens_fig:rmcube}, north of \object{NGC~1265}, is also a
good bubble candidate. It has a slightly curved shape. Its angular
size is approximately $30\arcmin\times15\arcmin$, which corresponds to
$750\times 375\ \mbox{kpc}$ at the distance of the cluster. It is
striking that it is located just north of the steep spectrum tail of
\object{NGC~1265} and has roughly the same curvature. As can be seen
in Fig.~\ref{brentjens_fig:blob1265}, its Faraday depth is
approximately 50--60~rad m$^{-2}$. This is close to the value observed
in the main part of the tail by \citet{ODeaOwen1986}.  Could this
structure be a detached bubble associated with a previous phase of
activity of \object{NGC~1265}? The projected separation of about
15$\arcmin$ translates to about 300 kpc or 300 million years for a
transverse velocity of 1000~km~sec$^{-1}$. We have also searched for
polarized emission in the steep spectrum tail of
\object{NGC~1265} \citep{SijbringDeBruyn1998}, but did not find any
significant ($\ga$~1~mJy~beam$^{-1}$) peaks.

The bar structure still further north is rather straight, but is 
both narrower and shorter than the two fronts on the west side of 
the cluster. Such structures have been discussed by
\citet{EnsslinBrueggen2002}.  With the current limited  
information on the detailed properties of these regions it is 
inappropriate to start making a detailed comparison.

\section{Remaining puzzles and future work}
\label{brentjens_sec:puzzles_and_future}

Through the discovery of highly polarized, very extended and very low
surface brightness emission associated with the Perseus cluster we
believe we have uncovered a rich diagnostic tool. This tool has the
potential to study the magneto-ionic medium, the relativistic plasma
content, the magnetic field topology and strength and the interaction
with LSS-formation related inflow into clusters of galaxies.
\citet{KeshetEtAl2004} predict that cosmological shock waves should be
easily detectable with the Square Kilometer Array (SKA) and the Low
Frequency Array (LOFAR) at frequencies below 500~MHz. They
suggest that the signal could already be marginally detectable with
current instruments and they are probably correct.

In future papers, following the reduction and analysis of several
recently acquired new full polarization datasets on the Perseus and
other clusters, at wavelengths from 21~cm to 250~cm, we will address
some of these issues in considerably more detail. The detection of
similar structures in other clusters would support our current
hypothesis for their origin. In order to minimize complications due
to high Faraday depth foreground polarization confusion, these
clusters are located at high Galactic latitude. We have chosen the
\object{Coma} and  \object{Abell~2256} clusters. Recently, filamentary
structures have been observed in the \object{Abell~2255} cluster
\citep{GovoniEtAl2005}. The  surface brightness of the
structures found in \object{Abell~2255} is about 1 to 2 orders of
magnitude higher than the structures we found in or near the Perseus
cluster. We expect that using RM-synthesis one could trace
these structures well into the large-scale filaments of the cosmic
web.

On the observational side we can identify the following tasks that lie
ahead of us:
\begin{itemize}
\item Image the corresponding total intensity and determine
the spectral properties of the emission.
\item Determine the polarization percentage and structure over 
an even wider range of frequencies.  This may allow us to 
unravel the internal plasma density and intrinsic  magnetic field structure.
\item Determine where the boundaries of the interaction 
between cluster and  large-scale structure flows in the intergalactic 
medium are located. 
\item Extend the sample of clusters
\item Combine the current data with data taken at different observing
frequencies.
\end{itemize}
The last point is necessary in order to derive the magnetic field
direction in the plane of the sky. For this one needs the polarization
angle at $\lambda = 0$, $\chi_0$. Unfortunately the $1\sigma$ error on the
derived Faraday depth is too large (more than $1\ \mbox{rad}\
\mbox{m}^{-2}$) to allow derotation to $\lambda = 0$. The $3\sigma$
error in position angle due to a $1\sigma$ error of $1\ \mbox{rad}\
\mbox{m}^{-2}$ at $\lambda_0^2 \approx 0.77$ is $2.3\
\mbox{rad}$. This implies that the 99\% confidence range is more than
4/3 times the $180\degr$ range that the position angle can assume.

On the theoretical side we are faced with the following questions:
\begin{itemize}
\item How can we explain the uniformity of the excess RM, in both sign
and magnitude, across the face of the Perseus cluster? 
\item How can we distinguish between the various models 
for the origin of the relativistic particles at the edges of the
cluster (are they from re-energized relics or are they shock-accelerated?)
\item How long can AGN bubbles hibernate before they fail to be 
re-energized to radiate at observable radio frequencies?
\item What are the observable effects if massive galaxies, or
galaxy sub-groups, plough through such bubbles? The numerical 
simulations suggest that this would happen rather frequently.
\end{itemize}

Bubbles carry important information on the final phases of radio 
galaxies and the study of AGN fossils may bring us somewhat closer 
to an answer on the long-standing question: do radio galaxies stop
being active rather suddenly or do they just fade away?
The study of bubbles and relics may quantify the role 
played by AGN activity and its feedback on cluster evolution. 
Cluster evolution and radio AGN activity are clearly linked 
in the central parts of clusters and may provide the answer 
to 'the cooling catastrophe' 
\citep{FabianEtAl2003ShocksAndRipples,ClarkeEtAl2004}.
Are enough bubbles blown to
prevent such a catastrophe?  It is believed that AGN activity is recurrent
\citep{SchoenmakersEtAl2000} but it is at this moment unclear how many 
bubbles exist in a cluster, nor do we know how long they can
hibernate. Is the large-scale polarized structure that arches around 
the steep spectrum tail of \object{NGC~1265} indeed 
the remains of an earlier  phase of activity?


\section{Conclusions and summary}

\label{brentjens_sec:conclusions_and_summary}

This paper has described low frequency observations of the Perseus
cluster obtained with the WSRT. The total intensity data agree with
earlier observations and do not present new insights into the cluster
emission. However, using a novel method of wide band polarimetric
imaging, called RM-synthesis, we have been able to extract 
information on extremely faint polarized signals buried deep in the
noise of individual narrow band frequency images.  

Polarized emission has been detected in the Perseus cluster over a
wide range of Faraday depths from about 0 to 90~rad~m$^{-2}$.  Low
$\phi$ emission ($\phi <$~15~rad~m$^{-2}$) is attributed to the local
Galactic foreground.  Emission at values of $\phi >$~30~rad~m$^{-2}$
on the other hand, shows organized structures on scales up to a degree
and displays rapidly fluctuating polarization angles on scales of the
order of $1\arcmin$--2$\arcmin$.  A Galactic foreground interpretation
for the high $\phi$ emission appears extremely implausible.  The
diffuse structures have a clear excess Faraday depth of about
+40~rad~m$^{-2}$ relative to a dozen distant radio galaxies
surrounding the Perseus cluster and are probably located at the near
side periphery of the cluster.

Most of the polarized emission, which is located at distances of
$0\fdg 5$--$1\fdg 5$ from the cluster centre, appears 1--2 orders of
magnitude too bright to be explainable as Thomson scattered emission
of the central radio source off the thermal electrons in the cluster
\citep{Syunyaev1982}.  However, this remains a viable explanation for
previously detected highly polarized 21~cm emission from the inner
$10\arcmin$--$20\arcmin$ as well as part of the 81--95~cm emission
observed in the region surrounding \object{NGC~1275}.  We believe that
the bulk of the emission associated with the Perseus cluster is
related to buoyant bubbles of relativistic plasma, probably relics
from still active or now dormant AGN within the cluster. A lenticular
shaped structure, referred to as the lens, and measuring 0.5--1~Mpc is
strikingly similar to the structures predicted by
\citet{EnsslinGopalKrishna2001} and may be in the process of being
re-energized.  A bright, about 0.4~Mpc sized, structure located to the
north of the steep spectrum tail of \object{NGC~1265} may be evidence
of an even earlier phase of activity from this famous head-tail radio
galaxy \citep{SijbringDeBruyn1998}.

We have not yet detected the total intensity corresponding to any of
these polarized structures. This may be due to a combination of
dynamic range in Stokes $I$ and lack of short spacing
sensitivity. However, there is no doubt that the structures are very
highly polarized ($>$20\%), as is common for many relic sources in
clusters.

At the western edge of the cluster we have detected linear structures
of several Mpc length that may be related to shocks caused by infall
of gas into the Perseus cluster along the Perseus-Pisces filamentary
structure of the cosmic web \citep{EnsslinEtAl1998, Burns1998}.

The ability of low-frequency RM-synthesis to image very extended
polarized structures, at intensity levels far below the total
intensity image which may be limited by confusion or dynamic range
problems, makes it a very powerful technique.  Regions of low surface
brightness and low magneto-ionic plasma density are predicted to occur
at the periphery of clusters and extending into the inter cluster
medium and the interface with regions shocked in the process of
large-scale structure formation.  The radio spectra associated with
such emission are often (predicted to be) steep requiring observations
at low frequencies.  The Low Frequency Array (LOFAR, see
http://www.lofar.org) is the ideal instrument to embark on such
studies.  It combines excellent surface brightness sensitivity and a
wide frequency range with the capability to perform high angular
resolution studies in cases where beam depolarization is important.

\begin{acknowledgements}
The Westerbork Synthesis Radio Telescope is operated by ASTRON
(Netherlands Foundation for Research in Astronomy) with support from
the Netherlands Foundation for Scientific Research (NWO). The
Wisconsin H-Alpha Mapper is funded by the National Science Foundation.
\end{acknowledgements}


\bibliographystyle{aa}

\bibliography{2992}

\end{document}